# Looking for an out: Affordances, uncertainty and collision avoidance behavior of human drivers

Leif Johnson[1*], Johan Engström[1], Aravinda Ramakrishnan Srinivasan[2,3], Ibrahim Özturk[2], Gustav Markkula[2]

## Abstract

Understanding collision avoidance behavior, that is, how human road users respond to urgent traffic conflicts, is of key importance in traffic safety research and for designing and evaluating advanced driver assistance systems (ADAS) and autonomous vehicles (AVs). While existing experimental work has primarily focused on response timing in traffic conflicts, the goal of the present study was to gain a better understanding of human evasive maneuver decisions and execution in collision avoidance scenarios. To this end, we designed a driving simulator study where participants were exposed to one of three surprising opposite direction lateral incursion (ODLI) scenario variants. The scenario variants differed in terms of the incursion trajectory of the encroaching vehicle, thus generating different kinematically available escape paths for the participants. The results demonstrated that both the participants' collision avoidance behavior patterns and the collision outcome was strongly determined by the scenario kinematics and, more specifically, by the uncertainty associated with the oncoming vehicle's future trajectory. In scenarios where a relatively certain escape path is available human drivers reliably exploit it to avoid collision. By contrast, when facing a scenario with high uncertainty about available escape paths, human drivers are less decisive and typically fail to avoid collision. We discuss pitfalls related to hindsight bias when judging the quality of evasive maneuvers in uncertain situations and suggest that the availability of escape paths in collision avoidance scenarios can be usefully understood based on the notion of *affordances*, that is, agent-relative opportunities for evasive action, and further demonstrate how such affordances can be operationalized in terms of reachable sets. We conclude by discussing how these results can be used to inform computational models of collision avoidance behavior.

---

[1] Waymo LLC, 1600 Amphitheatre Parkway, Mountain View, California 94043 USA
*Corresponding author: leif@waymo.com
[2] University of Leeds, Woodhouse Lane, Leeds, West Yorkshire LS2 9JT United Kingdom
[3] Chair and Institute of Highway Engineering, RWTH Aachen University, Mies-van-der-Rohe-Straße 1, 52074 Aachen

# Introduction

Understanding how human road users respond to urgent traffic conflicts—such as a lead vehicle braking, a pedestrian suddenly entering the road ahead, or a vehicle encroaching into an oncoming traffic lane—is of key importance in traffic safety research and, more specifically, for designing and evaluating advanced driver assistance systems (ADAS) and autonomous vehicles (AVs) (Markkula et al., 2012; McDonald et al., 2019; Engström et al., 2024). Such responses, which we henceforth refer to as *collision avoidance behavior*, involve assessing the threat and finding a feasible escape path that avoids collision by braking, steering, and/or accelerating. While there is a relatively large body of literature on the timing of initial responses to traffic conflicts (e.g., Dozza, 2013; Green, Markkula et al., 2016; Engström et al., 2024; Olson, 1989; Summala, 2000), less is known about how drivers decide on and execute evasive maneuvers in collision avoidance scenarios.

Some studies have looked at evasive maneuvers of human drivers in real world crashes based on crash reconstruction and/or data from onboard Electronic Data Recorders (Kaplan and Prato, 2012; Riexinger, Sherony and Gabler, 2019). These studies report widely different findings, even with respect to whether drivers exhibited any evasive responses at all (32% in Kaplan and Prato, vs. 100% in Riexinger et al.). Such differences are most likely due to differences in data fidelity and point to the challenges of using accident reconstruction or EDR data to study human collision avoidance behavior. For this reason, most work in this area has focused on experimental studies in driving simulators and, to some extent, naturalistic driving data.

Markkula et al. (2016) analyzed naturalistic rear-end (front-to-rear) crashes and near crashes from the SHRP2 naturalistic dataset (Hankey, Perez and McClafferty 2016). The results showed that (1), in line with Riexinger et al. (2019), drivers exhibited an evasive maneuver in the great majority of the events and (2) the evasive maneuver magnitude (final steady state longitudinal acceleration and jerk) increased with scenario urgency. However, this study did not analyze the choice of what evasive maneuvering to apply (e.g., braking versus steering). This was addressed by Sarkar et al. (2021), also using naturalistic rear-end conflicts from SHRP2, who fitted statistical models to identify key parameters determining drivers' evasive maneuver (braking / swerving) choice. One general finding was that a decision to steer to avoid collision depends on the availability of a kinematically available escape path. However, the notion of a kinematically available escape path was not operationalized in this study. The analysis also indicated that drivers tend to prefer braking to steering in urgent situations, where time to collision is already low when the driver becomes aware of the hazard.

Several experimental studies on human driver collision avoidance behavior using driving simulators have systematically investigated the impact of scenario-related factors on evasive maneuver behavior. These studies have typically focused on lateral incursions in straight crossing path (SCP) scenarios at intersections. In one early driving simulator study, Lechner and Malaterre (1991) investigated collision avoidance behavior in a SCP scenario with an initially stopped *principal other vehicle* (POV) encroaching from the right and then stopping in the path of the *subject vehicle* (SV) driven by the study participant. The study implemented three different scenario variants with times-to-collision (TTC) of 2.0, 2.4, and 2.8 seconds. TTC was here defined as the projected time from the start of the POV's incursion until the SV reached the projected collision point, assuming the SV's current speed remained constant. The results showed that the likelihood of the driver braking only (without steering) increased with longer TTC (when the driver had more time available). Steering left in front of the POV (i.e., in the same direction as the POV's incursion) was mainly performed in scenarios with shorter TTCs. By contrast, opposite direction steering to the right, behind the POV, mainly occurred for longer TTCs. Moreover, when swerving behind the POV (opposite direction), the participants tended to brake first and then steer. By contrast, participants steering in front of the POV tended to swerve only and not brake at all. These results thus indicate that the participants adopted different evasive maneuver strategies depending on the kinematics of the scenario. Specifically, drivers preferred to brake, or steer behind the POV (i.e., opposite direction), when there was sufficient time available and tended to swerve in front of the POV (same direction) when there was less time available. The great majority (80%) of the participants collided with the POV, and the collision involvement was strongly related to the criticality of the situation, which depended both on the TTC and the speed adopted by the participant at the start of the incursion and operationalized in terms of the deceleration required to stop. This scenario urgency also appeared to influence the drivers' evasive maneuvering strategies in ways not fully predicted by TTC alone.

Lechner and Maleterre (1991) also analyzed response times to the incursion, where a distinction was made between the driver's “first identifiable action” and the “first active input”. The former included either accelerator release or steering while the latter category included steering or braking evasive responses. The general finding was that response times in this scenario were relatively fast: around 0.8 s for accelerator release and steering response and 1 s for braking response. TTC did not have any effect on “first identifiable action” response times but did significantly affect “first active input” response times. However, no further details or explanations of this latter effect were provided in the paper.

The general results reported by Lechner and Malaterre (1991) on how scenario kinematics influence evasive maneuver behavior in SCP incursions have been replicated in several more recent studies. Weber et al. (2015) studied evasive responses in an SCP incursion scenario using a Vehicle-in-the-Loop (VIL) driving simulator, where participants operated a real vehicle in a closed area while perceiving a virtual world displayed through a head mounted display. Participants were exposed to eight different SCP conflict scenarios with a POV encroaching from the right, where the scenarios differed with respect to the kinematics of the POV. In contrast to Lechner and Malaterre (1991), in all scenarios the POV continued across the intersection rather than braking to stop. The main finding was that, in line with Lechner and Malaterre (1991), participants tended to brake only or steer right (opposite direction) in scenarios with high TTC (around 4 s; TTC was here referred to as Time to Arrival, TTA) while steering left (same direction) in scenarios with low TTC (below 2 s).

Similar results were obtained in a driving simulator conducted by Hu and Li (2017). Here, the kinematics of an SCP scenario were varied in terms of *priority level* such that the SV and POV, assuming constant approach speeds, either reached the intersection at the same time (zero priority level), the SV arrived just before the POV (positive priority level), or the POV arrived just before the SV (negative priority level). Thus, a positive priority level corresponds to a higher TTC/TTA in Lechner and Malaterre (1991) and Weber et al. (2015) and vice versa. Similar to Weber et al. (2015), the POV continued across the intersection rather than braking to a stop (as in Lechner and Malaterre, 1991). While the previous studies only included incursions from the right, the study by Hu and Li (2017) also included incursions from the left. For the right incursion, the results were generally aligned with the previous studies: same direction (left) steering with less tendency to brake dominated for zero and positive priority levels (small TTC) while braking only and/or opposite direction steering dominated for scenarios with a negative priority level (high TTC). However, the pattern of responses for the left incursion was quite different, with a higher prevalence of opposite direction steering (60% compared to 35% for the right incursion) and a lower tendency to brake only for the negative priority level (high TTC) scenario. This last finding, which may possibly be related to the geometrical asymmetry between the right and left incursions (where the POV first appeared at a larger lateral distance in the left incursion), suggests that relatively small differences in scenario kinematics could have large effects on the evasive maneuvering response patterns.

While the studies cited so far focused on the SCP scenario, a driving simulator study by Li, Rakotonirainy and Yan (2019) investigated evasive maneuvering behavior in three different scenario types: (1) a right incursion SCP scenario (similar to the previous studies), (2) an encroaching pedestrian scenario and (3) an opposite direction lateral incursion (ODLI) scenario. However, this study did not include kinematic variations within each scenario type. In all three scenarios, the encroaching road user started from a stationary position and then moved at constant speed into the path of the SV. Overall, the participants' evasive maneuvering strategies differed strongly between the three scenarios. For the SCP scenario (referred to as right-angle collision in the paper) it was found that all participants chose to brake only, which, in line with the results above, was explained by a relatively long TTC (5 s) when the POV initiated the incursion. In the pedestrian scenario, where the pedestrian encroached into the SV's lane from the right, participants appeared to select two distinct strategies, either braking only or braking and swerving to the left (same direction). For the ODLI scenario (referred to as the head-on collision scenario in the paper), the majority of the participants (51%) braked only, while 16% braked and steered to the right (same direction), and 33% braked and steered left (opposite direction). Since the ODLI scenario was set up such that the POV continued to the right across the SV's lane, the participants who braked only or steered right (same direction) typically ended up in a collision while those who swerved left (opposite direction) all avoided collision. Interestingly, those who steered in the opposite direction had a significantly longer steering response time than those who swerved right (3.38 s vs. 1.41 s, average).

Taken together, the existing studies reviewed above suggest that human drivers' evasive behavior is strongly determined by scenario kinematics. In particular, for the right incursion SCP scenario, the studies suggest a consistent pattern of where the choice of evasive maneuver strategy (i.e., brake only, swerve only left/right or brake and swerve left/right) is largely determined by the time the SV has available when the POV initiates the incursion (operationalized in the studies by TTC/TTA or priority level). However, it should also be noted that this pattern was observed at the aggregate level; the individual variation was still large and different drivers often adopted different evasive strategies in the same scenario. While providing some important hints on how drivers' evasive strategies depend on scenario kinematics, the existing studies yield an incomplete picture in several ways.

First, most of the studies focused on a single scenario type, SCP, and mostly on right-hand incursions. For the left incursion scenario included in Hu and Li (2017), the consistent pattern of responses found for right incursions was significantly altered. Moreover, the finding from the SCP scenarios  that braking only is mainly preferred at longer TTCs seems to be at odds with the results of Sarkar at el. (2021) who found that, in rear-end conflicts, drivers preferred braking in urgent situations with low TTC. Thus, it is unclear to what extent the relatively consistent findings for the SCP right incursion scenario generalize to other types of conflict scenarios. As demonstrated analytically by Brännstrom, Coelingh and Sjöberg (2014), the kinematic availability of braking and swerving escape paths at a given TTC depends strongly on speed, such that escape paths involving braking are generally more kinematically available at lower speeds and escape paths involving swerving become more kinematically available at higher speeds. In line with this, Lechner and Malaterre (1991) found that the approach speed, which determines scenario criticality, also seemed to influence evasive maneuvering choices in the SCP scenario. Thus, there is still a lack of a general understanding of what aspects of scenario kinematics determine the evasive maneuvering choices of human drivers.

Second, existing studies typically have not considered drivers' perceived *uncertainty* about the future behavior of other road users. Rather, several of the studies (Hu & Li, 2015; Li et al., 2019; Weber et al., 2017) interpret steering maneuvers in the same direction as the POV as "irrational" behavior given that this often resulted in collisions in the implemented scenarios, whereas swerving in the opposite direction typically avoided collision. Weber et al. (2017) even refer to steering in the same direction as "swerving into danger". Li et al. (2019) further suggested that same direction steering is the result of the driver applying automated "heuristics" as opposed to an "algorithmic" approach considering all available options to make a more rational choice. However, this line of reasoning is subject to *hindsight bias* (Fischoff, 1975), where the appropriateness of a given evasive maneuver (in this case same direction steering) is judged *after the fact*. Many of the scenarios in Weber et al., (2017), Hu and Li (2015) and Li et al., (2019) were designed such that the POV continued through the intersection (or in the ODLI scenario in Li et al. crossed the SV's lane) and thus could be avoided by opposite direction steering. However, this may no longer be the case if the POV stops shortly after the incursion (as in Lechner and Malaterre, 1991), or if the oncoming vehicle in the ODLI scenario turns back into its own lane, thus leaving room for avoidance by same direction steering. Indeed, Lechner and Malaterre (1991) evaluated such counterfactual POV trajectories in their scenarios and showed that they significantly altered the collision outcome results. On the basis of this analysis, they conclude: "*This shows above all that the result of an emergency situation is completely uncertain, and that the behavior of the obstacle to be avoided, in particular in the case of an intersection, is a determining factor. The consequences of driver actions are therefore uncertain, even if some maneuvers have more chance of being successful than others.*" (p. 136). Thus, since the future trajectory of the POV is typically not known when the driver has to make their initial evasive maneuvering decision, the perceived uncertainty about the POV's future behavior is likely to be a key factor underlying evasive maneuvering decisions. However, this has not been systematically investigated in existing work.

Third, existing studies have typically focused only on the initial response to the conflict (e.g., braking combined with swerving). However, understanding human collision avoidance behavior requires accounting for the full sequence of driver responses. For example, the driver may initially attempt steering left but then reverse the decision and steer towards the right if an opportunity for avoiding collision opens up on the right side. While some existing studies indicate that the initial evasive response may differ from the maneuver eventually performed to avoid collision (McGehee et al., 1999), a detailed analysis of the sequence of evasive maneuvers in collision avoidance behavior has, to our knowledge, not yet been attempted in existing work.

Fourth, several of the studies cited above (Hu & Li, 2015; Li et al., 2019; Weber et al. 2017) exposed the participants to repeated critical scenarios. Thus, even if the order of the different scenarios was counterbalanced in the studies, it may be expected that, with repeated exposure, the participants' evasive responses are influenced by increased anticipation of critical events (Engström et al., 2010). While Hu and Li (2017) demonstrate that the approach speed was not affected by event repetition, and Weber et al., (2015) used a set speed to prevent anticipatory defensive driving strategies, it may still be expected that the evasive maneuvering behavior in repeated scenarios was influenced to some degree by anticipation. Indeed, the results in Weber et al. (2017) show that the collision rate in the first exposure was markedly higher than in the subsequent seven repetitions. In line with this, Lechner and Malaterre (1991) report that the response times for repeated trials were significantly shorter than for the first trial and, for this reason, they only included data from the first trial in their analysis of evasive maneuvering choices. Thus, the results from existing collision avoidance studies using repeated events should thus be interpreted with this issue in mind, and, to avoid this confounder, it is preferable to only expose each participant to a single critical event.

The general objective of the present study is to fill in some of the gaps on how human evasive maneuvering in critical collision avoidance scenarios depends on scenario kinematics, with the broad aim to develop a more generalizable account of human collision avoidance behavior. As suggested by McGehee et al. (1999), discussing the results of Malaterre et al. (1991), it appears that drivers encountering a traffic conflict attempt to "steer towards the biggest gap," or, more generally, seek available escape paths as determined by the kinematics of the situation, including their own evasive capabilities (Sarkar et al., 2021). We here suggest that such available escape paths may be usefully conceptualized as *affordances*, originally defined by Gibson (2014 / 1979) as what the environment "offers the animal, what it provides or furnishes, either for good or ill" (p. 127). In the present context, affordances can be seen as opportunities for evasive action to avoid collision, offered by the environment to the driver given the action capabilities (e.g., braking and swerving capacity on a given road surface) of the driver-vehicle system. However, to be useful for present purposes, affordances also need to account for the *uncertainty* about how the situation will play out, an issue that is not clearly addressed in traditional conceptualizations of affordances. In this paper, we propose a novel way to conceptualize, operationalize and visualize affordances for evasive actions based on reachability theory (Althoff & Dolan, 2014; Pek et al., 2020) in terms of non-collision states that are kinematically *reachable* given assumptions of one's own evasive capabilities and reasonably foreseeable behavior of other road users. We further discuss this somewhat broader notion of affordances in relation to traditional ecological psychology (Gibsonian) accounts at the end of the paper.

More specifically, we explore how different scenario kinematics influence evasive maneuver behavior in a scenario different from the SCP incursion scenario addressed in the existing literature, with the aim to shed light on underlying factors and mechanisms beyond TTC. To this end, we chose the ODLI scenario, where the POV unexpectedly enters into the SV's lane from an adjacent lane travelling in the opposite direction. This scenario was also chosen due to the high degree of uncertainty about the future path of the POV, which was hypothesized above as a key factor behind evasive maneuvering choices. This also allows for comparison with the results of Li et al. (2019) cited above, who, however, did not include scenario-specific kinematic variations.

We implemented three different kinematic variants of the ODLI scenario which differed in the “steepness” of the incursion, operationalized by incursion level (a concept derived from the priority level used Li & Hu, 2015). Here we analyze how the incursion level influenced the choice and performance of braking and steering evasive maneuvers. To avoid anticipation due to repeated events, each subject was only exposed to a single conflict scenario, similar to Lechner and Malaterre (1991).

Moreover, a key novel contribution of the present study is the analysis of *sequences* of evasive actions, thus addressing the closed-loop nature of collision avoidance behavior, potentially including replanning of evasive maneuvers as the scenario unfolds. We also conduct an analysis of reachable states to operationalize and visualize how kinematically available escape paths (i.e., affordances) are dynamically realized during the scenario, and how this relates to the observed human evasive maneuvering behavior.

# Method

## Experimental setup

We recruited 64 participants (Age M=41.25, SD=14.01, Min=22, Max=82) for the study. Of these, 40 were male and 24 were female. All participants were required to hold a valid UK driving licence for at least three years, drive at least once a week and have normal or corrected to normal vision. Due to technical and health and safety requirements, all participants were required to be between 152–198 cm tall, not be pregnant, not be taking any cardioactive medication, and not suffer from claustrophobia or severe motion sickness.

The study was advertised through the Virtuocity participant pool, social media channels and the University of Leeds mailing lists. Participants were asked to complete a form to indicate their eligibility and were then invited to book an experimental slot. All eligible participants received an informed consent form explaining the nature of the study and the driving simulator.

Before beginning the experiment, participants were given a trial run in the driving simulator to familiarize themselves with the controls (steering wheel, accelerator pedal, and brake pedal) and displays; note that the simulator and all participants used UK (left-hand) traffic rules. The participants were generally instructed to approach the driving task as they would in their normal, everyday driving, imagining that they would like to get to their destination without delay, and deal with speed limits and surrounding road users as they would normally do.

During the experiment, each participant drove the SV on a two-lane road through a suburban environment, with traffic in both directions. The participants were specifically instructed to maintain speed at the posted speed limit of 40 mph (65 km/h). A complete session took about 60 minutes, from participant arrival to departure, with a 15 minute main drive. At the end of the study, participants were compensated with £20.

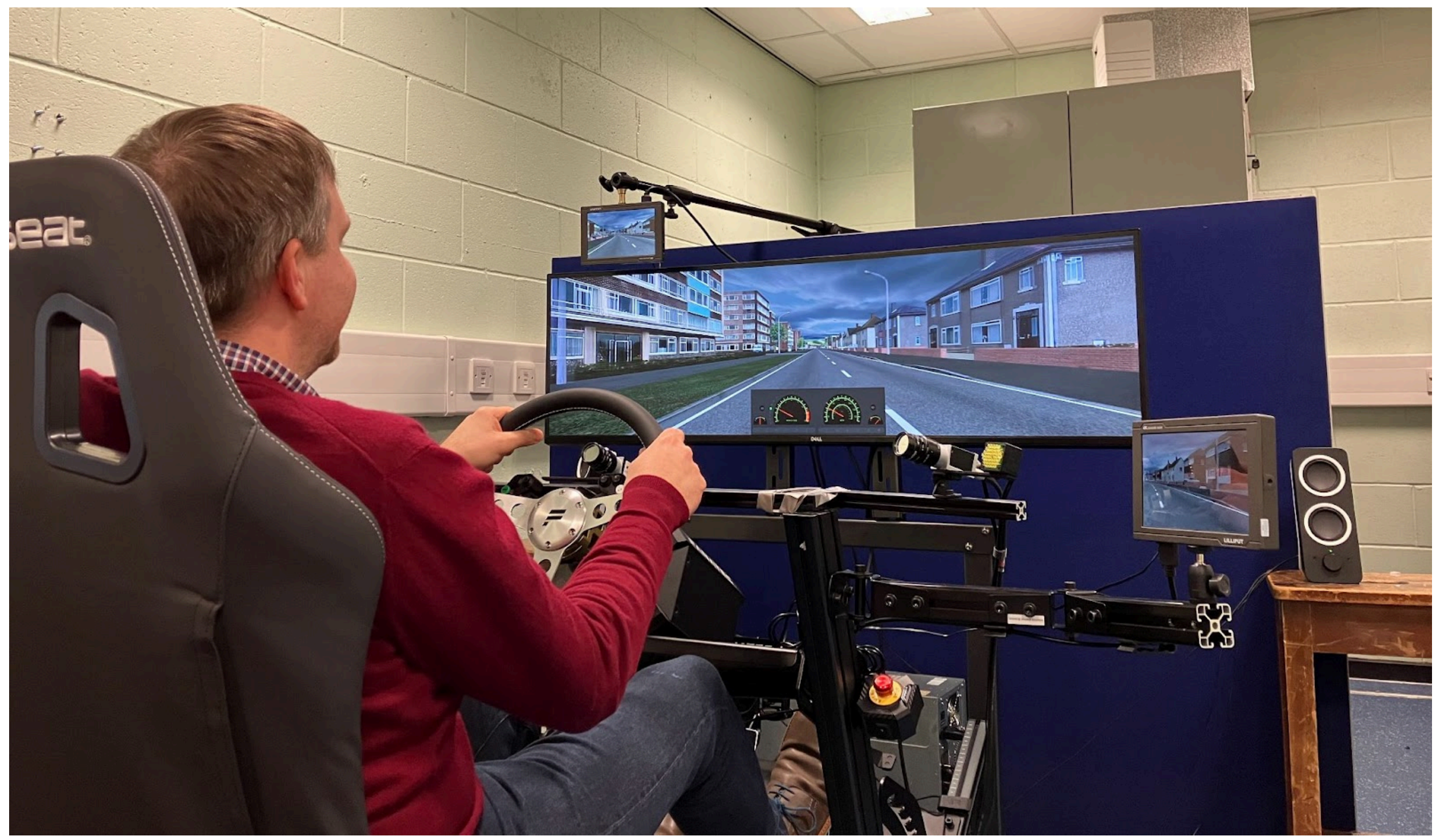

Figure 1. The Static Driving Simulator used in the present study.

The study was conducted using a fixed driving simulator with a high-resolution monitor, separate screens for left and right wing mirror and rear view mirror views (see Figure 1). The main screen was a 49-inch 32:9 monitor with a resolution of 3840 x 1080 pixels The movable seat, steering wheel, accelerator and brake pedals were mounted on a stable Next Level 130 Racing® Wheel Stand Direct Drive. The simulator software was developed in-house.

In the main drive, after the participant had performed approximately 13 minutes of normal (unsurprising) driving, one of the vehicles in the oncoming traffic lane (the POV) initiated an ODLI event. The POV was programmed to keep a constant speed of 40 mph, travelling in the right lane (from the perspective of the SV), in the opposite direction relative to the SV. When the two vehicles reached a 5.15 s longitudinal time gap, the POV initiated the ODLI event by departing from its lane and moving laterally along a fixed trajectory, modelled as a Bézier curve, towards the SV's lane, subsequently crossing the centerline of the road and entering the SV's lane. The POV's maneuver was timed to begin at the *trigger point* (henceforth referred to as $t_T$) so that a collision would occur at the *critical point* ($t_C = t_T + 5.15$ s) if the participant did not respond. The participant then had to perform an evasive maneuver (braking, steering, or a combination thereof) to avoid a head-on collision. We additionally defined $t_P$ as the time when the longitudinal proximity between the SV and POV reached zero (i.e., the vehicles either crashed or passed each other). Note that if a participant did not respond, then $t_P = t_C$ by design, but a participant's evasive maneuvers could change the timing of closest proximity, for example, by applying the brakes.

The experiment ended a few seconds after $t_P$, that is, after the SV passed the POV longitudinally, or when a collision occurred. In the latter case, the simulator screen was grayed out; otherwise the participant was instructed to bring the SV to a halt. To avoid behavioral biases related to anticipation of critical events (such as response priming or increased vigilance), each participant was only exposed to a single ODLI conflict scenario. The conflict scenario was varied by three incursion levels—*steep*, *medium* or *shallow* incursion—further described below. The incursion level, which determined the kinematic availability of an escape path (i.e., the presence of an affordance), was thus the independent variable in a between-groups experimental design. We targeted 20 participants in each group, with some over-recruitment to guard against data loss. In the end, no participants needed to be excluded, so the medium incursion group included 22 participants while the steep and shallow groups included 21 participants each.

## Incursion variations

We implemented three different variants of the ODLI scenario with different kinematic properties, operationalized as *incursion levels*, which is based on the "priority level" concept used by Li and Hu (2015). In general, the incursion levels differed with respect to the incursion trajectory of the POV, as illustrated in Figure 2, implementing different affordances for evasive maneuvers to avoid collision. In the first scenario variant, the *steep incursion*, the POV entered the SV's lane and continued crossing the lane to the SV's left, thus creating an affordance for steering towards the right (opposite direction). In the second variant, the *medium incursion*, the POV entered into the middle of the SV's lane heading directly towards the SV, leaving it more uncertain whether it would continue straight or go back to its original lane. Thus, in the medium incursion scenario, there was no clear affordance for steering either left or right, although collision could kinematically, in hindsight, be avoided by steering right towards the road center without braking. Finally, in the *shallow incursion* scenario, the POV remained near the center of the road, to the right of the SV, creating an affordance for steering slightly to the left (same direction).

Quantitatively, we defined the incursion level (IL) as the lateral position of the POV in the participant's lane at time $t_C$ (i.e., the point when collision would happen, if no evasive maneuver was taken by the participant in the SV). The POV's position was measured relative to a reference point centered laterally in the POV, 20cm forward from the longitudinal center of the vehicle (see Figure 3). Incursion levels were defined such that an IL value of 0 means that the POV's reference point was centered in the SV's lane at $t_C$; an IL value of -1 means that the POV's reference point reached the left lane boundary (i.e., the edge of the road) at $t_C$; and an IL of 1 means that the POV's reference point was laterally centered at the right lane boundary (i.e., the center of the road). In this study, we used IL values of -0.8 for the steep incursion, 0 for the medium incursion, and 0.9 for the shallow incursion. Figure 4 illustrates the POV incursion trajectories for the three scenario variants.

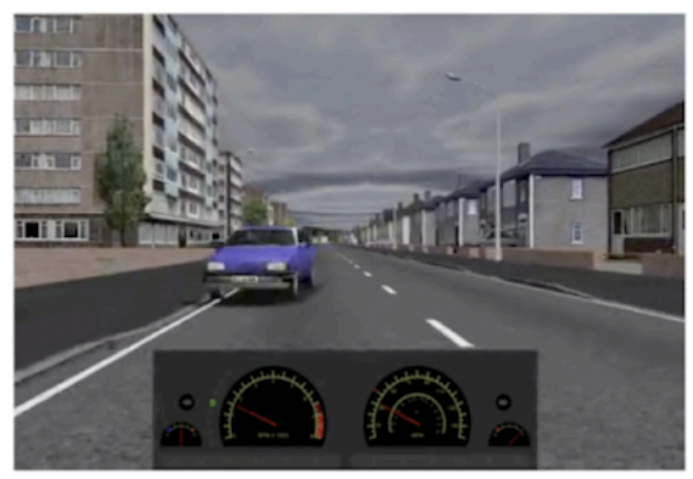
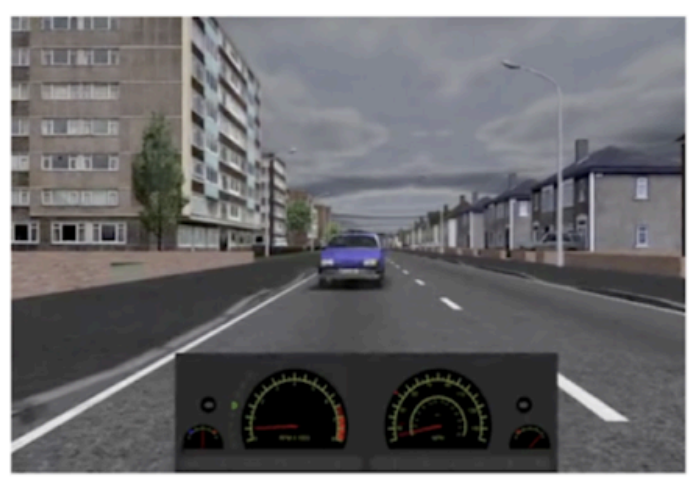
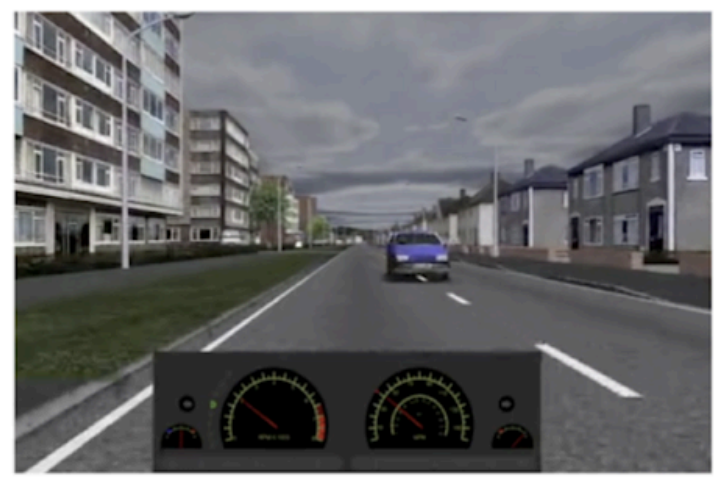

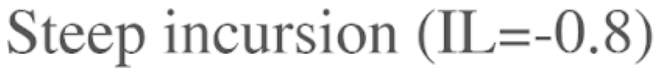

Medium incursion (IL=0)

Shallow incursion (IL=0.9)

Figure 2. Illustration of the three scenario variants (incursion levels, IL).

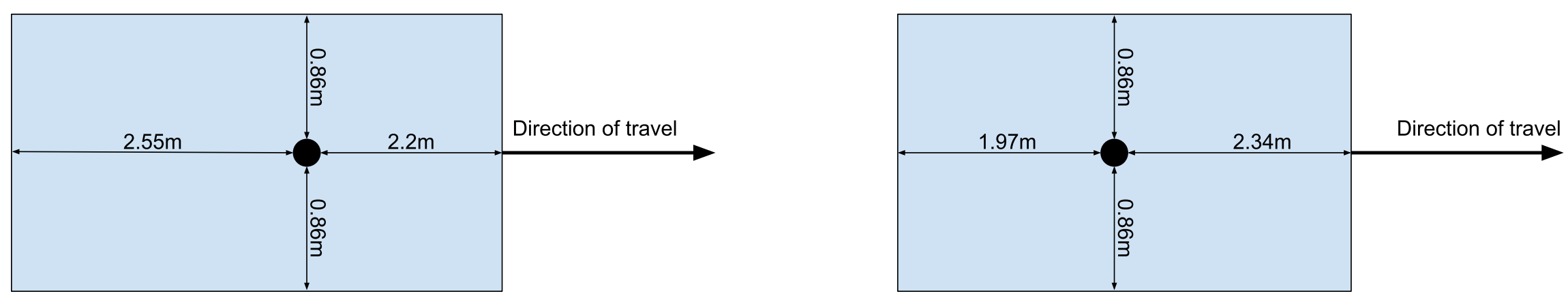

Figure 3. Dimensions and positioning reference point of the POV (left) and the SV (right).

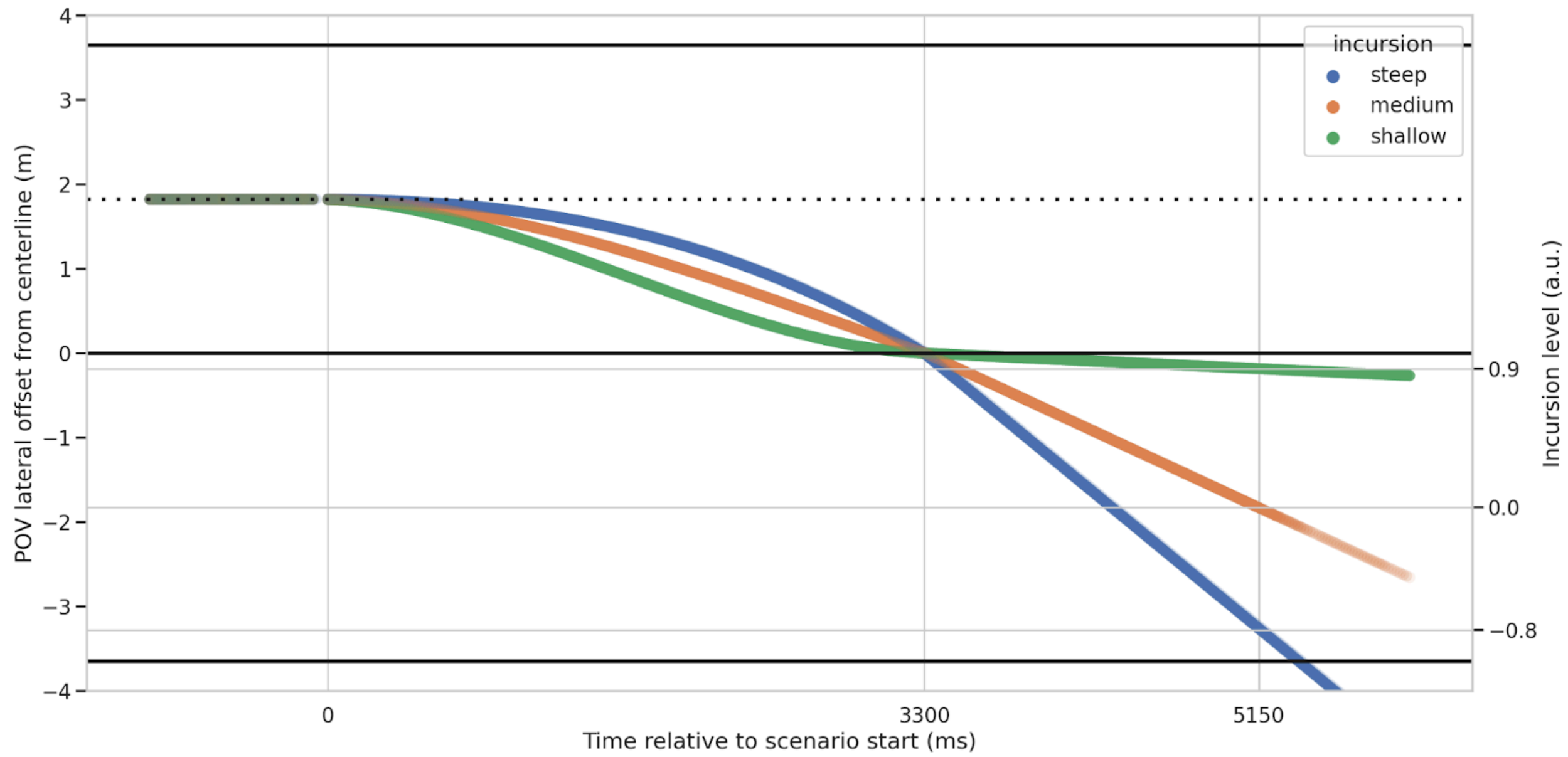

Figure 4. POV trajectories for the three scenario variants.

## Dependent measures

During the experiment, we gathered information about the collision avoidance responses displayed by participants. Analysis of participants' responses focused on a specific time window as follows. The end of the time window, $t_E$, was set equal to the time of closest proximity between the two vehicles, i.e., $t_E = t_P$. The beginning of the time window, $t_B$, was set 400 ms after the onset of the POV's lane change maneuver, i.e., $t_B = t_T + 400$ ms. The 400 ms delay prevented the analysis window from including participant actions that were not initiated as a result of the POV's movement, i.e., actions that happened before a minimal typical "reaction time" from the onset of the oncoming movement.

Within the time window between $t_B$ and $t_E$, we used participants' control inputs to the simulated SV to measure responses to the oncoming POV. The simulator was equipped with a brake pedal, an accelerator pedal, and a steering wheel, each with a continuous range of inputs: the accelerator and brake pedals varied from 0% (not pressed) to 100% (completely pressed), and the steering wheel reported positive and negative angles of deviation from the neutral (straight) position in degrees. We applied thresholds to these continuous-valued inputs to convert them to binary responses as follows:

- *Accelerator release:* When the accelerator pedal was pressed less than 3%, corresponding to an acceleration value of approximately 0 $m/s^2$.
- *Brake onset:* When the brake pedal was pressed more than 15%, corresponding to a deceleration value of 1 $m/s^2$.
- *Steering onset (left, towards shoulder):* When the steering wheel was turned 5 or more degrees to the left. The steering wheel angle threshold was set to a value that reliably captured steering maneuver onsets while not being impacted by noise in the steering wheel angle signal.
- *Steering onset (right, towards road center):* When the steering wheel was turned 5 or more degrees to the right.

Thus, we operationally defined each response as an event where the signal in question crosses a threshold in the relevant direction (e.g., the accelerator pedal is released below 3% or the brake pedal is pressed more than 15%). The participant's response time to the conflict was computed separately for each of these binary response actions. For each action type, we defined the response time as the time difference between stimulus onset $t_T$ and the first response of that type; e.g., for brake onset, the brake onset response time was $min_i ( t_i ) - t_T$, where $t_i$ are all the brake responses in the analysis time window between $t_B$ and $t_E$. Note that for response time calculations, any subsequent responses of the same type do not affect the metric; that is, only the first response of each type was used for computing the response time.

Finally, we also classified each participant's evasive behavior in terms of the outcome of the scenario. To identify outcomes, we examined the lateral distance between the SV and the POV at the time of closest longitudinal approach, $t_P$. Three outcomes were possible:

- *Collision:* Any lateral spatial overlap between the SV and the POV based on the vehicle dimensions illustrated in Figure 3.
- *Pass via center:* The SV passed the POV on the right side by steering towards the centerline of the road.
- *Pass via shoulder:* The SV passed the POV on the left side by steering towards the shoulder of the road.

## Reachability analysis

In addition to the data analyses described above, we also operationalized and visualized the affordances in the different ODLI scenarios, by applying reachability analysis on the kinematic traces produced by each participant. Briefly, reachability analysis computes a sequence of *reachable sets* representing possible future states of the world (Althoff & Dolan, 2014; Pek et al., 2020). Each reachable set $R_t(\tau)$ contains the kinematic states that are possible to attain at future time $t + \tau$, starting from the current state of the world at a given time $t$. $R_t(0)$ contains just the kinematic state (e.g., location, speed, heading, etc.) of the road user at time $t$. Subsequent reachable sets are, in principle, formed by applying all available kinematic controls to each reachable kinematic state from the previous time step. That is, $R_t(\tau + \Delta\tau)$ contains all kinematic states from $R_t(\tau)$, projected through a kinematic model having time gradient $f'$, using the available controls $U$ at $\tau$, for a duration of $\Delta\tau$:

$$R_t(\tau + \Delta\tau) = \{ s + \Delta\tau \cdot f'(s, u) : s \in R_t(\tau), u \in U(\tau) \}$$

where $s$ and $u$ are elements from the reachable-state and control-value sets, respectively. The reachability computation halts at a fixed time horizon, $\tau = H$, here set to 4 s.

When rolling out reachable states for a given future time step, the POV expands its reachable set first, and then the SV expands its reachable set. However, any reachable states for the SV that overlap reachable states for the POV are considered to be potential collisions. These overlapping states are pruned from the SV's reachable set. Following Pek et al. (2020), we refer to this pruned reachable set for the SV as its *drivable area*. The drivable area represents states that the SV can achieve without exposing itself to collision risk.

The reachable set thus depends critically on kinematic assumptions (e.g., maximum lateral and longitudinal accelerations) regarding reasonably foreseeable behavior of the POV and the evasive capacities of the SV ; see Table 1. The kinematic assumptions about the POV can be seen as a bound on the uncertainty regarding its future behavior. In addition, we include assumptions that the POV, prior to the incursion, will comply with established traffic norms and road rules (i.e., *normative expectations*, see Fraade-Blanar et al., 2025). In the current ODLI scenario, this mainly involves the assumption that the POV will remain in its lane until the incursion shows evidence to the contrary. Furthermore, we introduce separate lateral acceleration limits for the POV's left and right steering actions, so that its estimated future reachability is biased toward steering to the left (i.e., toward its original lane).

| Parameter | SV | POV |
|---|---|---|
| Speed (m/s) | 20 | 20 |
| Acceleration ($m/s^2$) | 5 | 5 |
| Deceleration ($m/s^2$) | 8 | 8 |
| Left Lateral Acceleration ($m/s^2$) | 6 | **4** |
| Right Lateral Acceleration ($m/s^2$) | 6 | **0** |
| Forward Jerk ($m/s^3$) | 10 | 10 |
| Backward Jerk ($m/s^3$) | 30 | 30 |
| Lateral Jerk ($m/s^3$) | 30 | 30 |

Table 1. Kinematic parameter limits for the SV and POV during reachability analysis. Both vehicles assume a jerk-limited kinematic bicycle model. Most parameters were identical for both vehicles, except for the lateral acceleration limits, highlighted in bold.

# Results

## Prevalence of response actions

Figure 5 shows the prevalence of the different response actions during the analysis time window. Since the incursion groups contained different numbers of participants, the prevalence is given as the number of responses divided by the number of participants in each group (converted to percentages). For a single event, multiple responses of different types could occur, but each response type was only counted once. For example, a participant might release the accelerator pedal, apply the brake pedal, and steer the SV toward the shoulder in the course of their response; this sequence of actions would add a count of one to each of the accelerator, brake, and steer-shoulder actions in the plot. It should be noted that the responses counted here do not necessarily correspond to the evasive maneuvers finally attempted by the participant. For example, the participant may initially have steered left (shoulder) and then steered right (center) to avoid collision; in such cases, both steering responses are counted in this analysis.

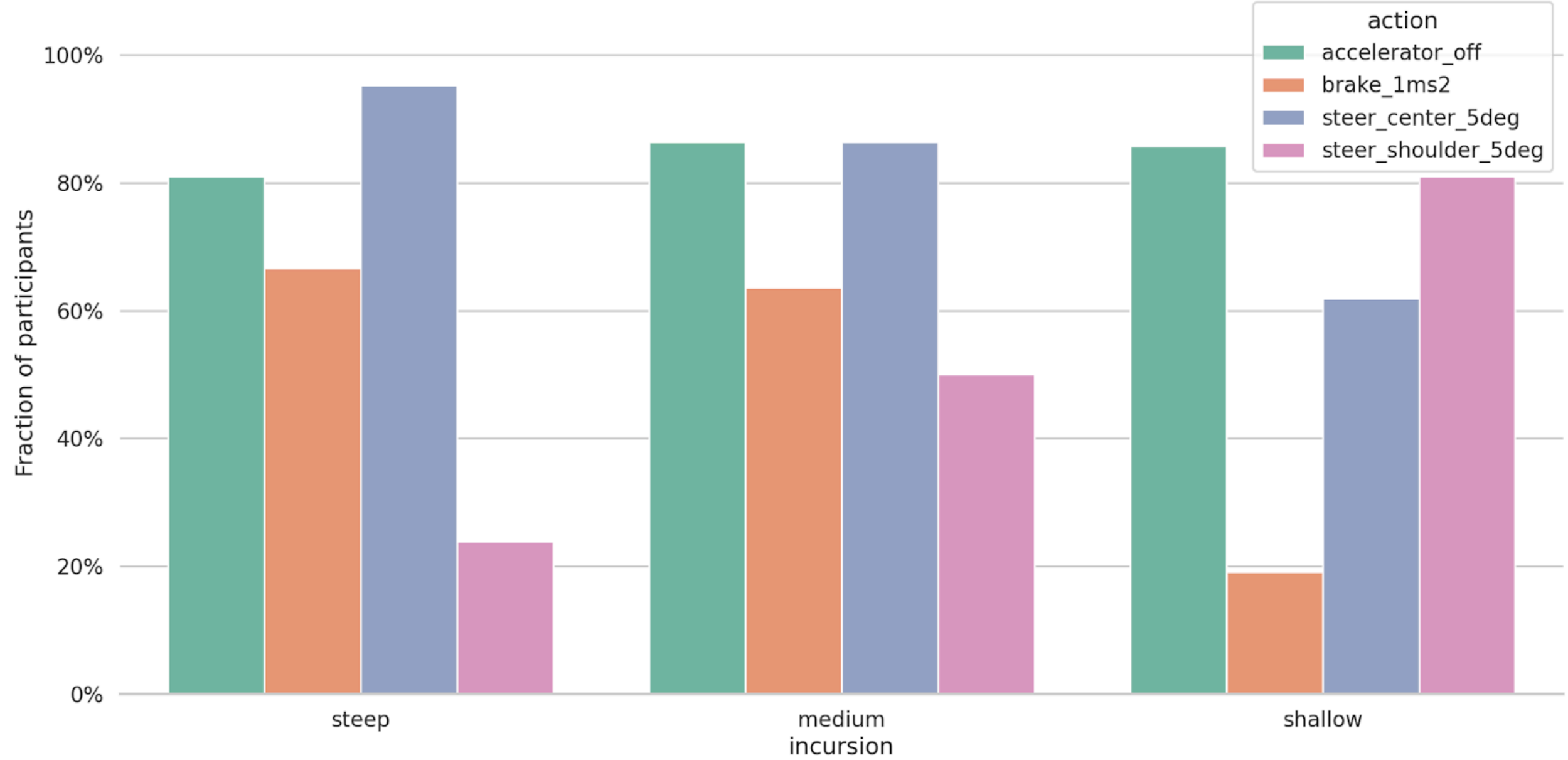

Figure 5. Fraction of participants showing accelerator, braking and steering responses for the three incursion levels.

As can be seen in the plot, the majority of the participants released the accelerator pedal in all three scenarios, with no significant differences between the incursion scenario groups ($\chi^2$ = 0.044, p = 0.978). Of the 10 cases that did not involve an accelerator pedal release during the analysis window, the participants had either already released the pedal prior to $t_B$ (in 9 cases) or kept the accelerator pressed down while simultaneously pressing the brake pedal (one case).

For the steep and medium incursion scenario the majority (around 60%) of the participants braked. This contrasts with the shallow incursion scenario where only around 20% of the participants braked (while, as just noted, most participants had still released the accelerator pedal). This is likely due to the fact that in the shallow incursion scenario, collision was typically avoidable by a relatively minor steering maneuver to the left (towards the shoulder), which did not necessitate braking. Braking behavior across incursion levels differed slightly ($\chi^2$ = 6.01, p = 0.050).

Finally, most participants attempted some form of steering in all three scenario variants, where the direction of steering depended strongly on the incursion level for steering towards the shoulder ($\chi^2$ = 6.67, p = 0.036) while the effect of incursion level was less for center steering ($\chi^2$ = 1.54, p = 0.462). In the steep incursion scenario, all participants but one steered right towards the road center, while only a few (around 20%) steered left towards the shoulder. In the medium scenario, participants exhibited steering in both directions with the majority (around 85%) steering to the right towards the road center and about 50% of the participants steering left towards the shoulder (recall that a single participant may have exhibited both steering actions in a sequence, as further analyzed below). In the shallow scenario most of the participants (about 80%) steered left towards the shoulder and a majority (about 60%) also steered right towards the road center. Further analysis revealed that right steering actions in the shallow incursion scenario mainly involved corrective steering to stay on the road that occurred within the analysis window (while most of the corrective actions in the steep and medium scenarios occurred after the end of the analysis window $t_E$). Further analysis also revealed that a few subjects (3) in the shallow scenario did not exhibit a steering response above the threshold but still avoided collision, since they had initially positioned themselves so far left in the lane that no further steering was required to avoid collision.

To summarize the results on response prevalence, the incursion level did not influence the accelerator release response (almost all subjects released the accelerator pedal after $t_E$) but influenced the prevalence of both the braking and the steering left/right responses. In the steep and medium scenarios most subjects braked while only a minority braked in the shallow scenario. In the steep incursion scenario, steering responses, when exhibited, were always towards the right while the medium and shallow incursion scenario included both left and right steering, with left steering dominating for shallow incursions and right steering for the medium scenario. The right steering in the shallow incursion scenario mostly involved corrective steering (after the initial evasive left steering maneuver) while right steering in the steep incursion scenario typically constituted the initial evasive maneuver.

## Outcome

While the response prevalence results reported in the previous section indicate that the choice of steering evasive responses depends on the scenario kinematics (incursion level), this analysis does not reveal whether or not the evasive action was successful in avoiding collision and, if so, whether the participant eventually chose to pass on the right or left of the POV. Figure 6 shows the outcome frequency in the three scenario variants (again presented as percentages of the total number of participants in each group). As can be seen, it is clear that the incursion level played a key role in determining both if and how collision was avoided by the participants. In the steep incursion scenario, the great majority of the participants (90.5%; 19 of 21) successfully avoided collision by passing the POV on the right (i.e., towards the center of the road). Similarly, in the shallow scenario, most participants (95.2%; 20 of 21) avoided collision by passing the POV on the left side (i.e., between the POV and the curb). However, in the medium incursion scenario, almost all participants (86.4%; 19 of 22) ended up colliding with the POV. Crash outcomes were significantly different across the three incursion level groups, based on a chi-squared test comparing observed collision counts to expected counts using average collision rate ($\chi^2$ = 13.8, $p < 0.001$).

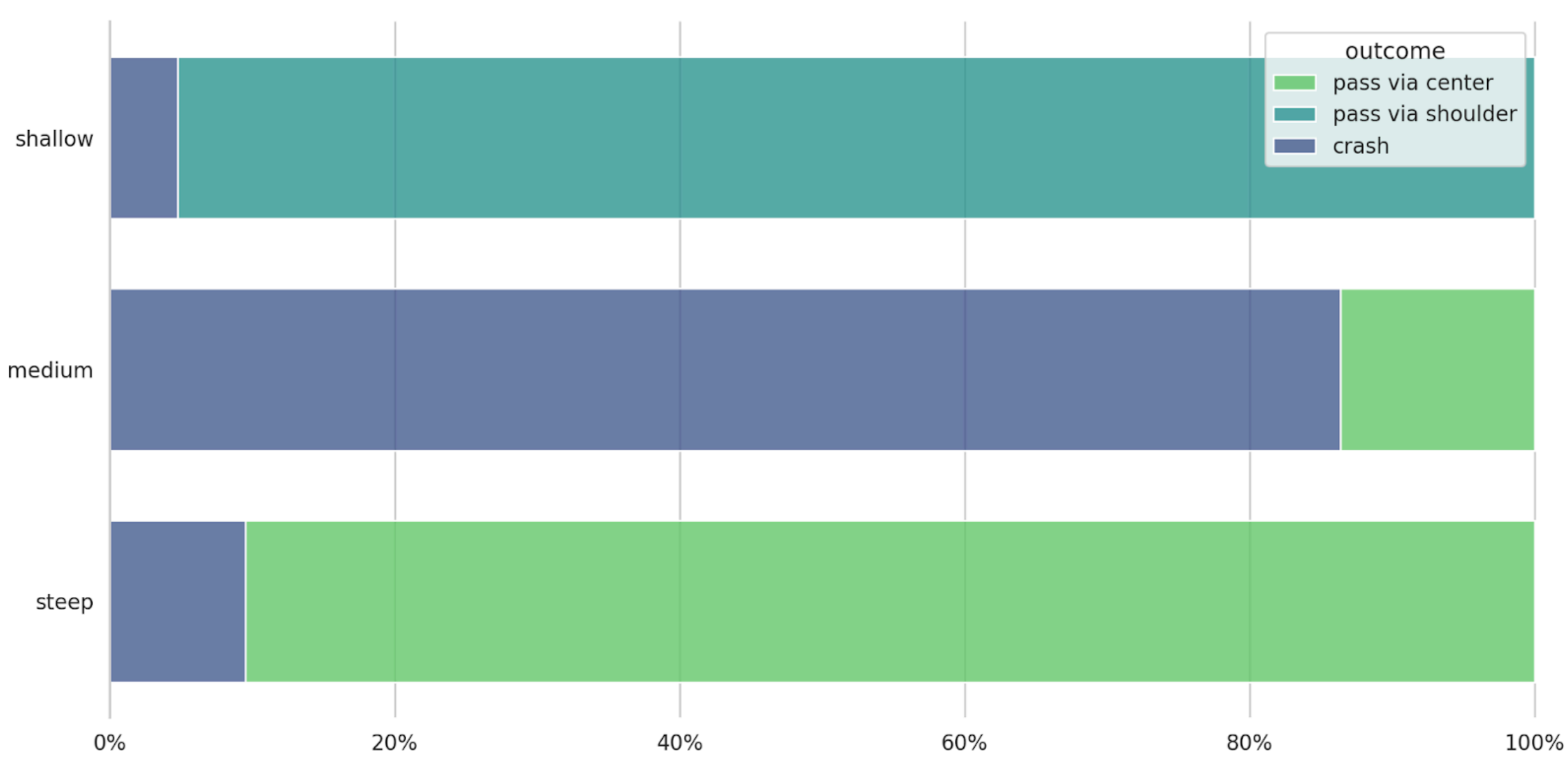

Figure 6. Distribution of scenario outcomes for the three incursion levels.

These results indicate that human drivers are able to perceive affordances (viable escape paths) when they exist, here in the steep and shallow incursion scenarios, and reliably maneuver in a way that avoids collision. By contrast, in the medium incursion scenario only 3 of 22 participants (13.6%) were able to avoid collision. Further analysis revealed that the successful participants in this scenario avoided collision by steering towards the right (towards the center) without braking. Thus, even if collision was (in hindsight) kinematically avoidable by swerving to the right, only a few participants managed to find this particular escape path.

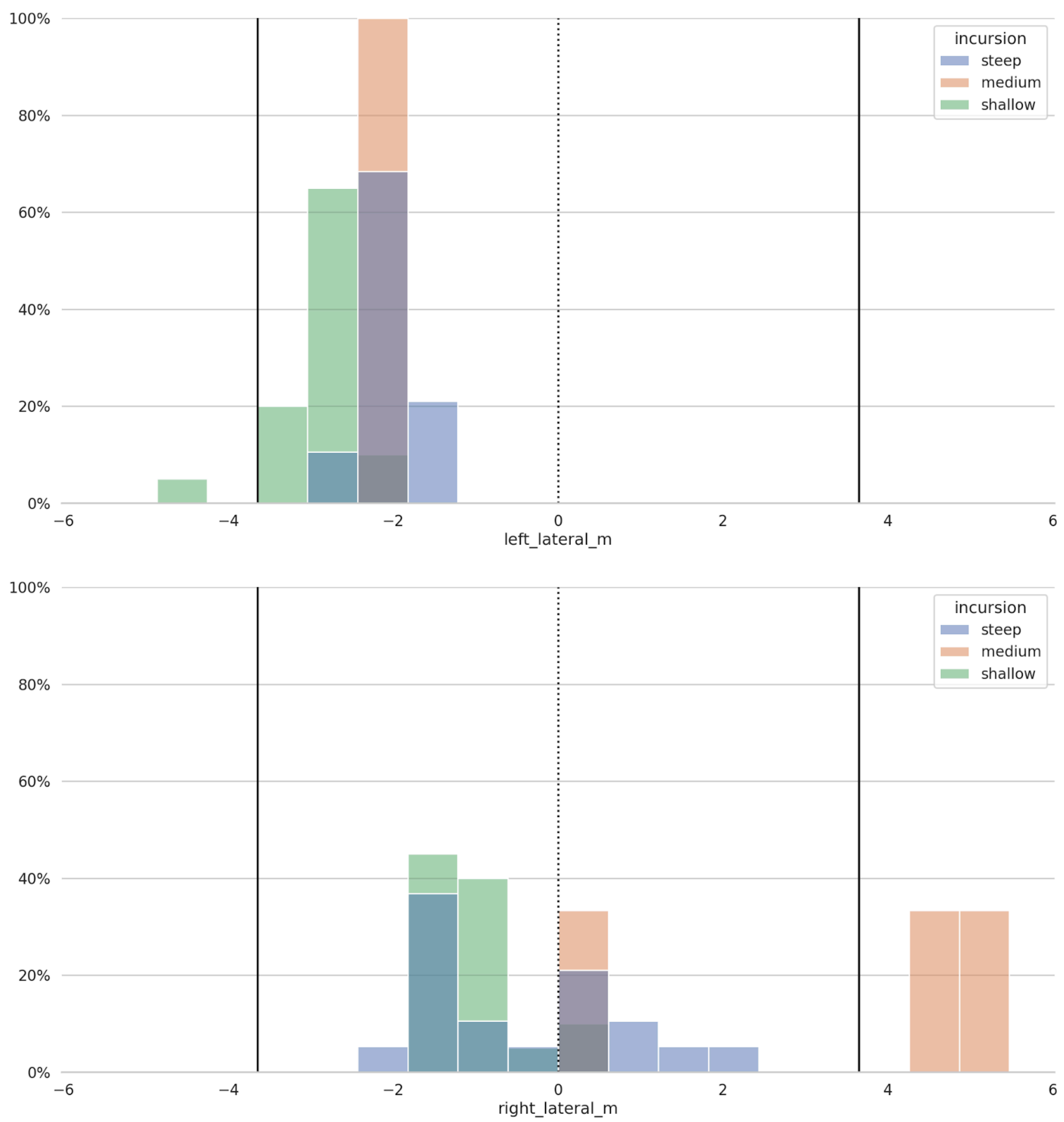

Figure 7. Distributions of lateral displacement extrema for participants who avoided crashing. The solid lines indicate the road edges and the dashed line indicates the road center. The top plot shows participants' maximal displacement to the left (negative lateral position relative to the road center), while the bottom plot shows participants' maximal displacement to the right (positive lateral position values). The solid lines represent the road boundary and the dotted line represents the centerline of the road.

We also analysed the extent to which participants who avoided collision were able to stay within the edges of the road after their evasive maneuver. The results are shown in Figure 7. A few participants, one in the shallow incursion group and two of the three participants who avoided collision in the medium incursion group, drove the centers of their vehicles past the edges of the road. In all cases this happened after the evasive maneuver, during corrective steering aiming to return to the original lane. Hence, it can be noted that only one participant in the medium incursion group managed to both avoid collision and stay on the road.

## Response sequence

To obtain a more detailed understanding of the participants' response processes in the different incursion scenarios, we analyzed the state transitions over time to observe sequential patterns in evasive response behaviors.

For each time step $t$ in the analysis window from $t_B$ to $t_E$, we converted the lateral control input at that time to one of three states $s_{lat}(t)$: steering toward the shoulder (left), no steering, and steering toward the center (right). Similarly, for each time step in the analysis window we converted the longitudinal control input to one of four states $s_{lon}(t)$: acceleration > -1 m/s$^2$ (cruising), acceleration between -1m/s$^2$ and -4 m/s$^2$ (soft braking), and acceleration < -4 m/s$^2$ (hard braking). The overall state at a given time $t$ was thus a pair of the lateral and longitudinal states: $s(t) = \{s_{lat}(t), s_{lon}(t)\}$. Then for each $t$ from $t_B$ to $t_E$, we recorded any state transition between $s(t\text{-}1)$ and $s(t)$. Note that we did not record self-transitions, that is, transitions where $s(t) = s(t\text{-}1)$; such transitions measure the amount of time that a participant remains in a given state, while for this analysis we were focused on the evolution of the states over time. A final state transition was added from $s(t_E)$ to one of three pseudo-states representing the collision avoidance outcome of the scenario; these final transitions permit analysis of the patterns of control states leading up to the critical time in the interaction. The results for the steep incursion are shown in Figure 8.

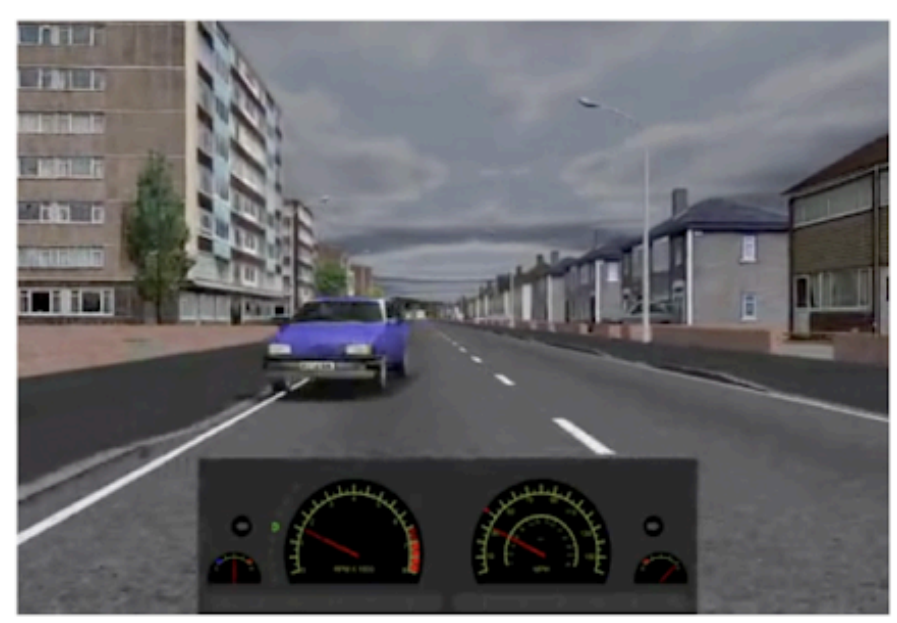

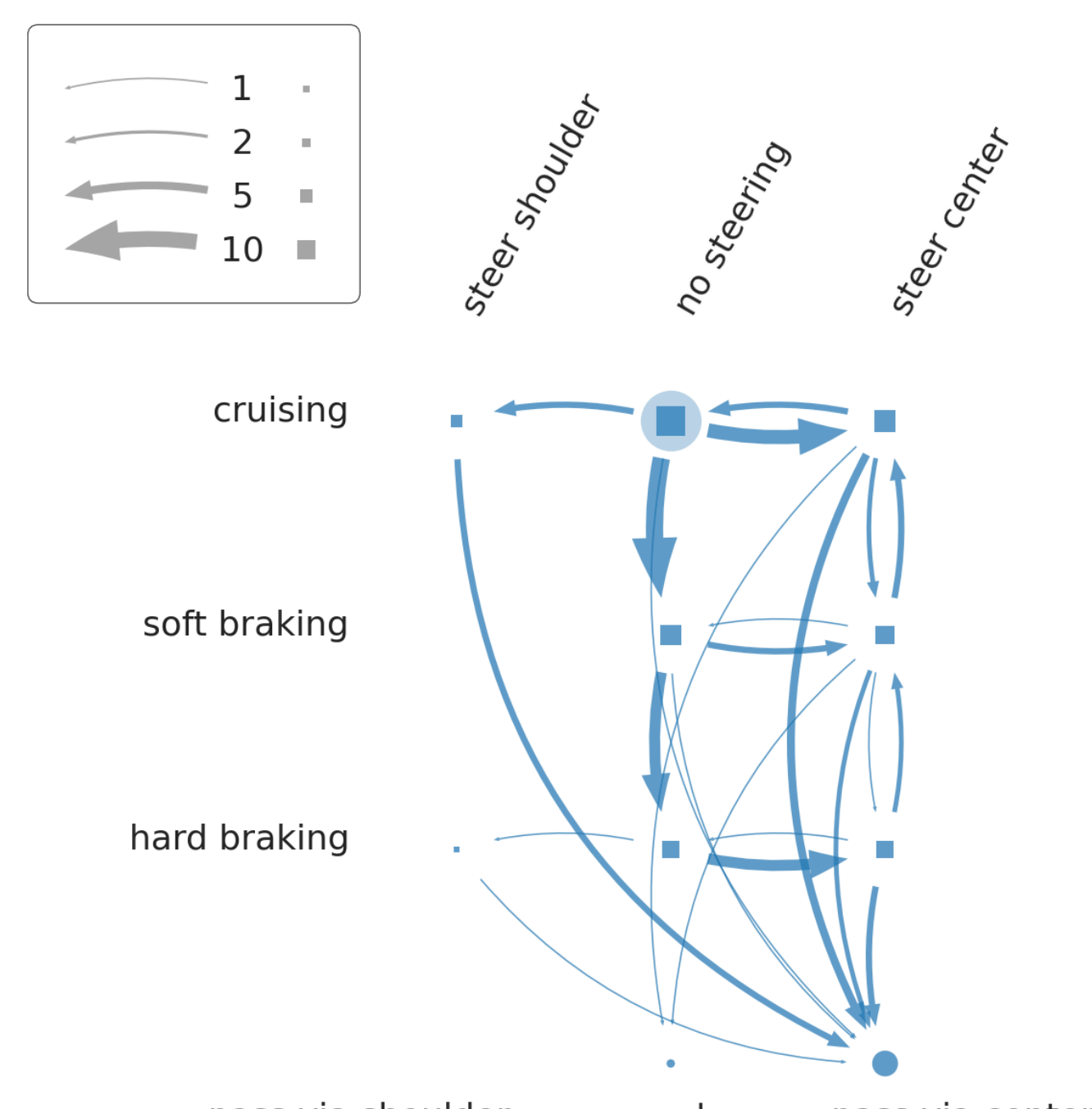

Figure 8. Sequence diagram for the steep incursion scenario. The ring indicates the initial states of participants at $t_B$. The thickness of the arrows and the sizes of squares indicate the count of observed transitions and state occupancies respectively.

As already shown by the response prevalence analysis above, evasive maneuvers in the steep incursion were dominated by braking and steering right (towards the center). The response sequence analysis in Figure 8 additionally reveals two main response patterns. The most common one is that the participant initially braked and then steered right and passed the POV via the road center. About half of these participants only braked softly (between 1 $m/s^2$ and 4 $m/s^2$) before steering right while the other half also braked harder (greater than 4 $m/s^2$). Further video inspection revealed that several of these participants essentially avoided collision by braking and then performing a minor steering maneuver to the right. The second main response pattern is that the participant steered only, without braking. Some of these participants, who steered early, likely initiated a corrective left steering maneuver before the end of the analysis window, which explains the path from *steer shoulder* to *pass via center*. It also can be noted that braking only, without any steering, was very uncommon. The few crashes represented a response pattern where the participants steered to the right (and braked softly in one case), but were still unable to avoid a collision. Manual inspection of these cases indicated that the steering response occurred too late to avoid collision.

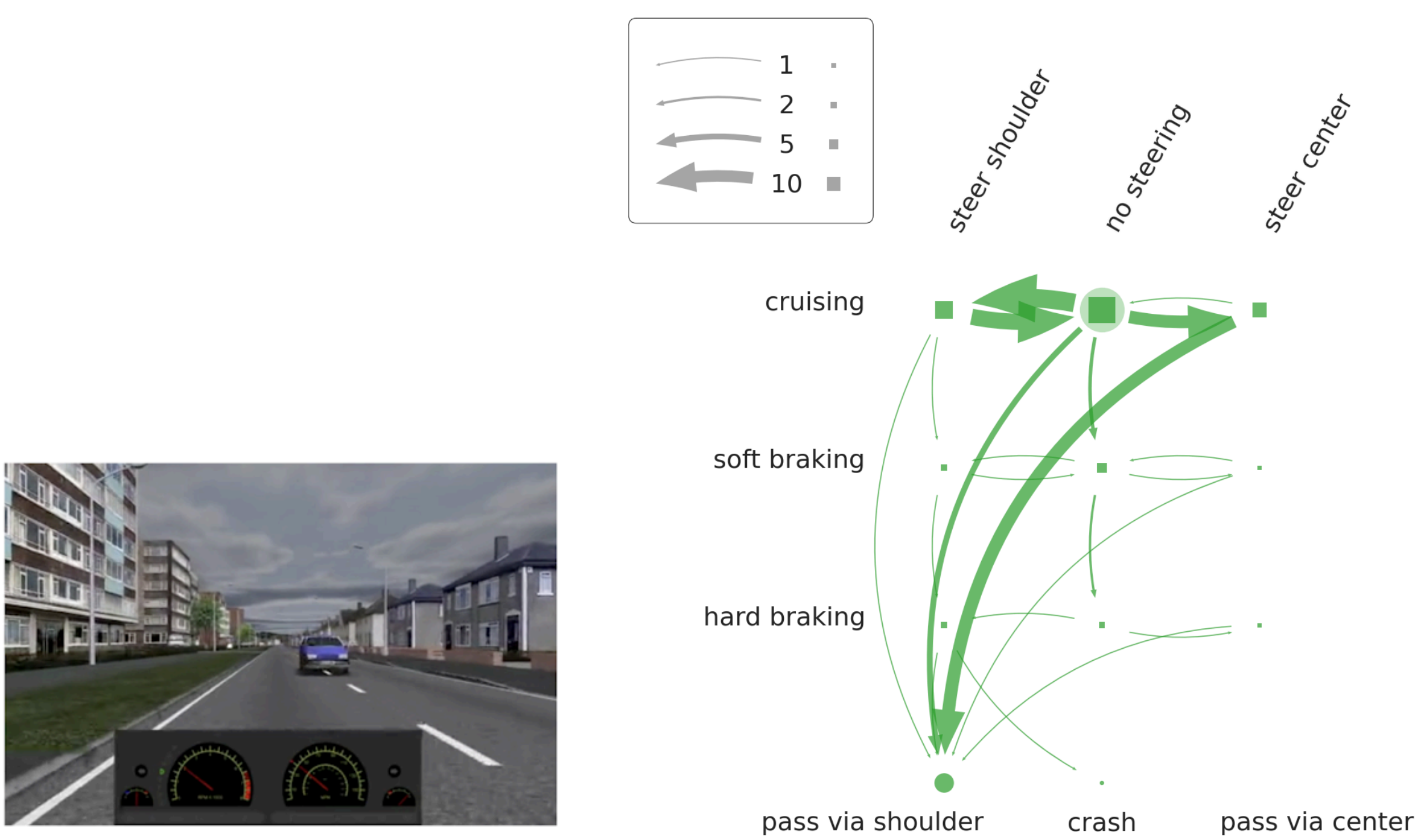

Figure 9. Sequence diagram for the shallow incursion scenario. Otherwise as Figure 8.

The sequence plot for the shallow incursion scenario is shown in Figure 9. The dominating response pattern is steering towards the shoulder (left) without any braking, followed by an early corrective steering maneuver to the right (center) which typically led to successfully avoiding collision by passing to the left of the POV (via shoulder). Since less steering is required here compared to the steep incursion scenario, the state sometimes returns to “no steering” within the analysis time window which explains the frequent “looping” path at the top. The path from “no steering” to “pass via shoulder” consists of cases where the participant was initially positioned so far left in the lane that no, or very minor (sub-threshold), steering was necessary to avoid collision. Overall this analysis confirms that in this scenario, the great majority of the participants avoided collision by performing the single evasive action of steering slightly to the left.

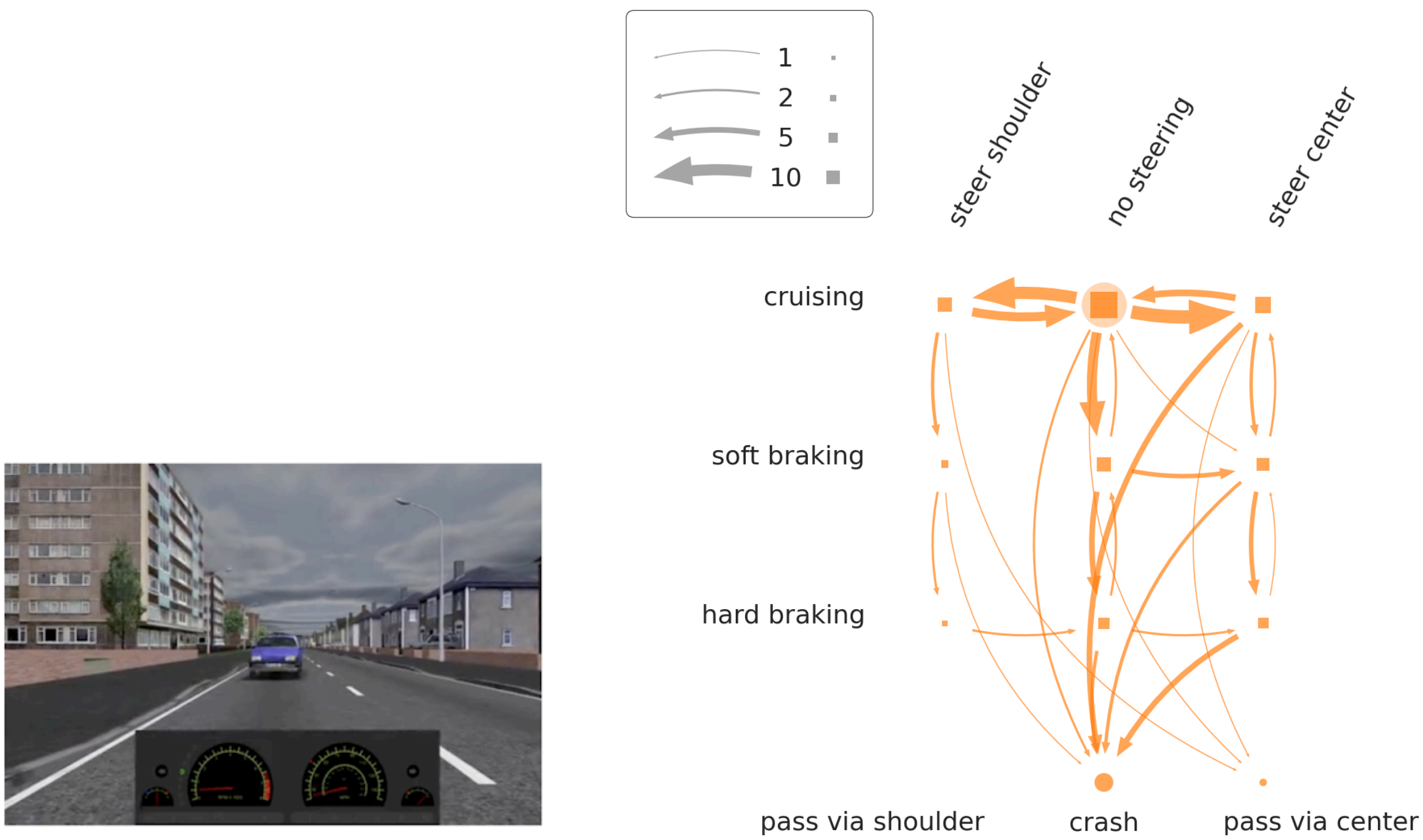

Figure 10. Sequence diagram for the medium incursion scenario. Otherwise as Figure 8.

The results for the medium incursion scenario are shown in Figure 10. Compared to the steep and shallow incursion scenarios, the response patterns exhibited by the participants are much more diverse. There is roughly equal prevalence of the first action being braking, steering right (towards the road center) or steering left (towards the shoulder respectively. A few participants attempted to avoid collision by braking only, but most steered right (center) at some point prior to the collision. Another group of participants steered right first and then braked, while yet others steered only. As noted earlier, the three participants who successfully avoided collision were all in this latter group.

There is also a relatively strong pattern of initially steering left towards the shoulder but then steering back again. This pattern is most prevalent before any braking begins and is completely lacking in the steep incursion scenario. This suggests that, for the medium incursion, many participants had an initial tendency to steer left (same direction) which was subsequently corrected by a larger steering action towards the right.

## Response timing

When analyzing response timing, a distinction can be made between the initial *reaction* to the event and the evasive *response*. Following Lechner and Malaterre (1991), the initial reaction was operationalized by the first response of any type, including the accelerator release, while the response was measured as the first response onset of braking or swerving.

The initial reaction times for the three incursion scenarios are shown in the left panel of Figure 11. An ANOVA revealed no significant effect of incursion level on initial reaction time (F = 0.44, p = 0.647). The response times for the three incursion scenarios are shown in the right panel of Figure 11.. An ANOVA revealed a significant effect of incursion level on response time (F = 4.3, p = 0.018), and a Tukey post hoc analysis revealed that only the difference between the steep and the shallow scenario was significantly different (q = 615.4, p = 0.014).

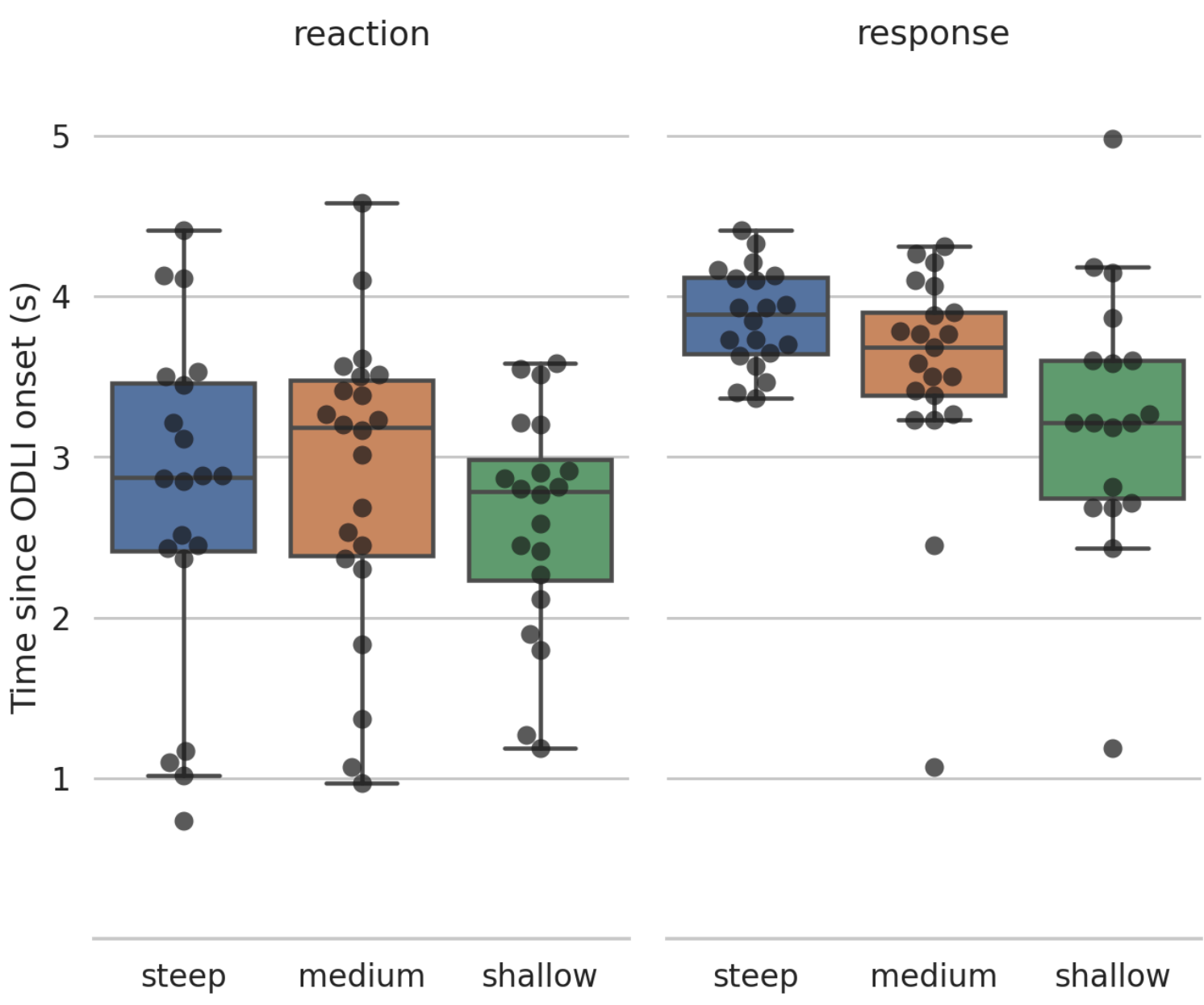

Figure 11. Initial reaction times (left) and evasive response times (right) for the three incursion levels. Lines on box plots show the 25th, 50th, and 75th percentiles, and whiskers show the most extreme data point within 150% of the interquartile range (difference between 75th and 25th percentile). Individual data points are shown using dots.

The response times for the individual response actions (accelerator release, steering and braking) are plotted in Figure 12. The results indicate that (as expected) the initial reactions in Figure 11 are dominated by accelerator release. Figure 12 also makes it clear that the effect of incursion level on response time in Figure 11 is mainly driven by differences in steering response time.

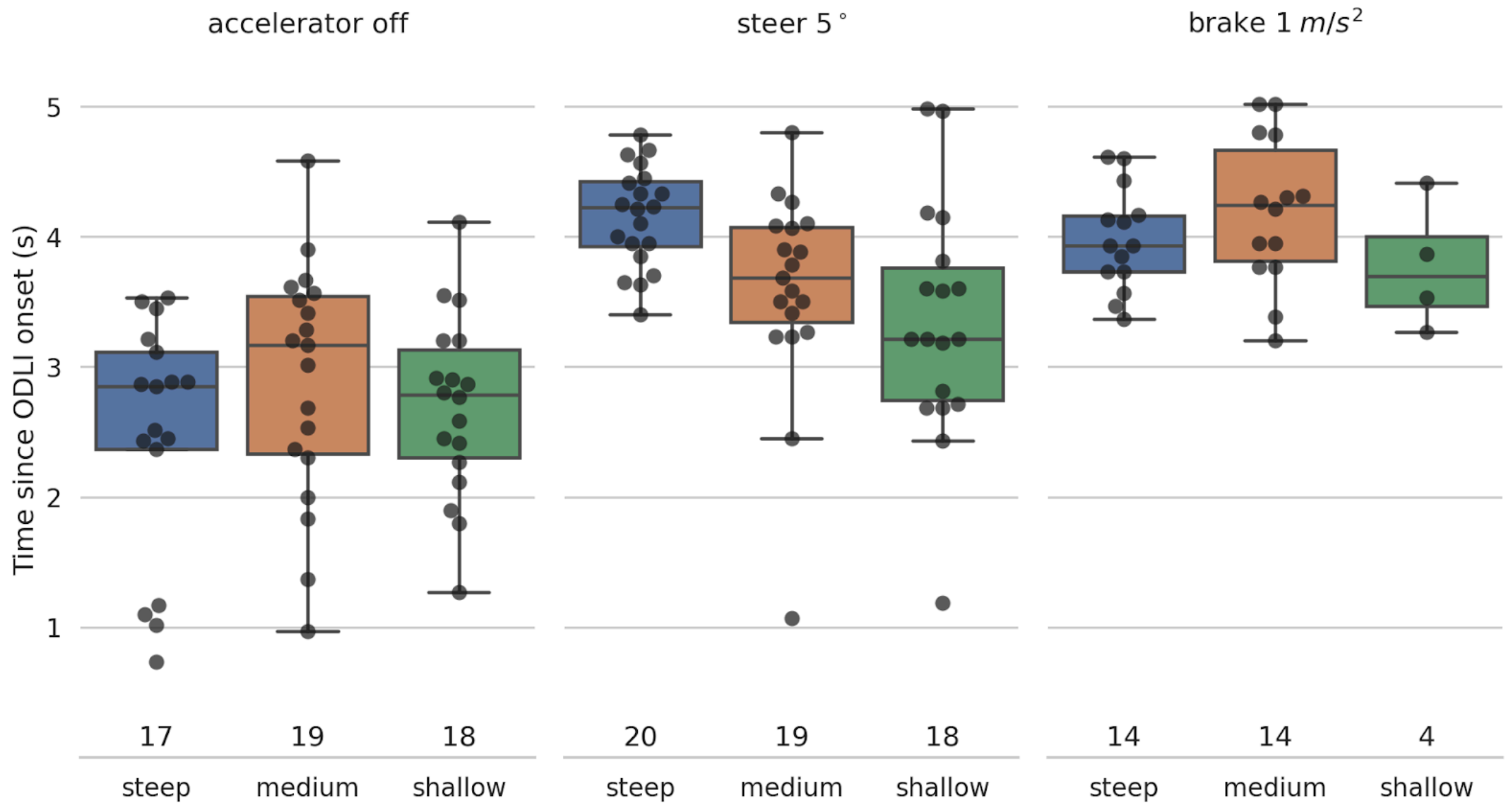

Figure 12. Response time for the different simulator control inputs. Boxes as in Figure 11. Numbers at the bottom of each plot indicate the number of data points in the boxes.

## Reachability analysis

Figures 13-15 show examples of how the reachable sets of the two vehicles evolve during the course of the scenario, for three individual participants. Figure 13 shows an example for a participant in the steep incursion scenario. Each plot shows reachable sets for one $t$ (current time) value, with the $t$ values increasing from left to right. After about t = 1 s, the POV has started its incursion into the SV's lane, so its reachable sets limit the safe drivable area for the SV; that is, the SV no longer has a drivable area for the full time horizon up to $\tau$ = 4 s when an affordance for steering right towards the center (opposite direction) opens up for the SV. This is because the POV's speed and assumed lateral acceleration limit (here set to 4 $m/s^2$, see Table 1) prevent it from returning to its original lane at that point. Note that the absence of a drivable area at any point in the interaction does not guarantee a collision—instead, the lack of a drivable area indicates that there is not a trajectory for the SV that is guaranteed to be collision-free given the kinematic assumptions defined in Table 1.

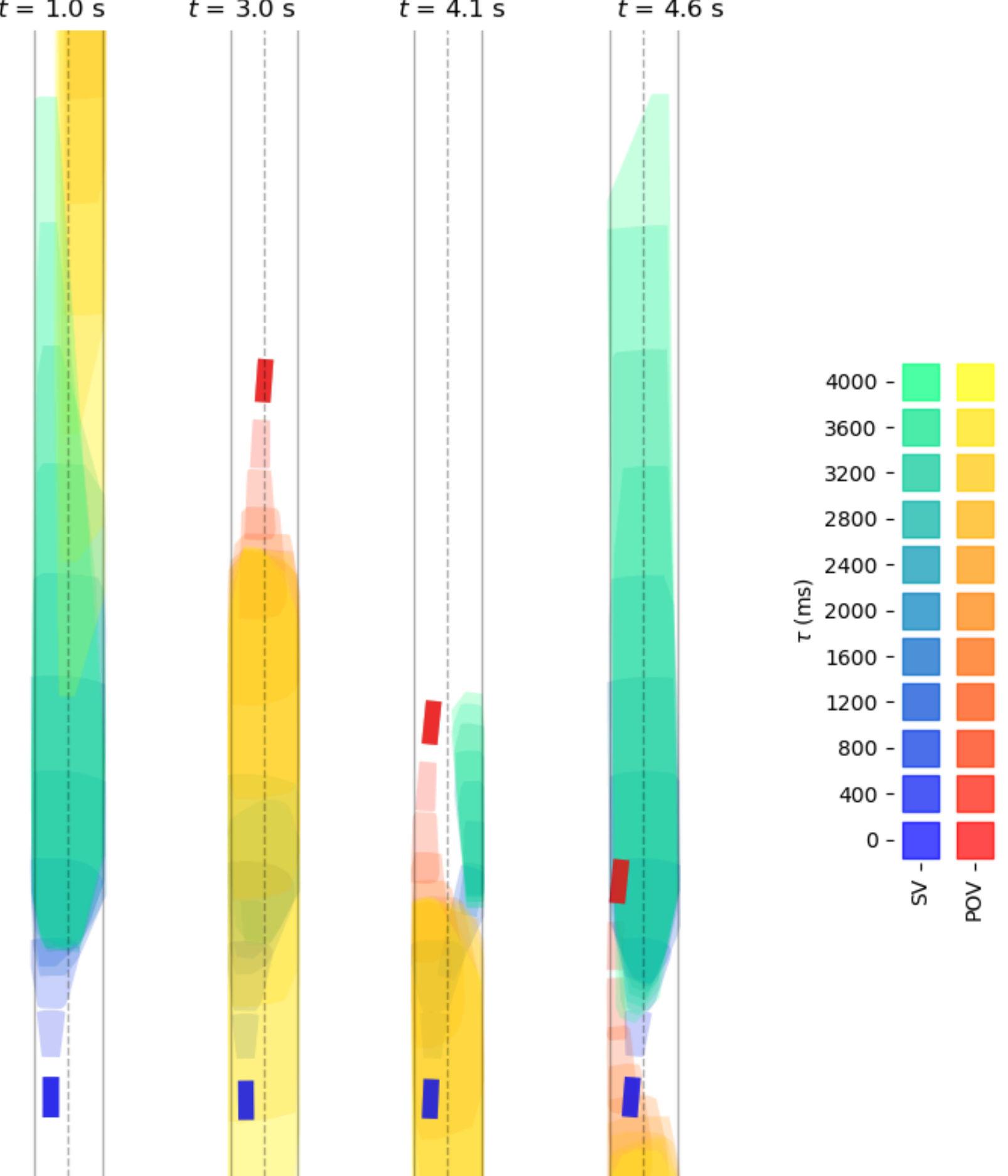

Figure 13. Illustration of reachability modeling for one participant in the steep incursion scenario. The SV is driving upwards, and the POV is driving downwards. Colors show reachable areas at different points in the future, given the initial state shown by the dark rectangles for the SV (blue) and POV (red). (The POV is outside the picture in the leftmost plot.)

Figure 14 shows an example reachability analysis for a participant in the medium incursion scenario. In this scenario, the SV loses its drivable area once the POV crosses into its lane, and because of the assumed kinematic capabilities of the POV, and the resulting uncertainty about its future trajectory, the SV never regains its drivable area. Thus, in the medium incursion scenario, there are no available escape paths in any direction due to the strong uncertainty about whether the POV will continue across or move back into its own lane, and a clear affordance never materializes. This ambiguity offers an explanation for the great individual variance in the participant's evasive behavior reported above for the medium incursion scenario, where several participants initially attempt to swerve left only to later reverse the decision and steer right. Only three participants, who all refrained from braking, were able to avoid collision by steering right, as discussed above. However, while this choice of evasive maneuver was here successful in hindsight, it would have resulted in collision if the POV had instead turned back into its own lane (and several of the participants that crashed would have successfully avoided collision).

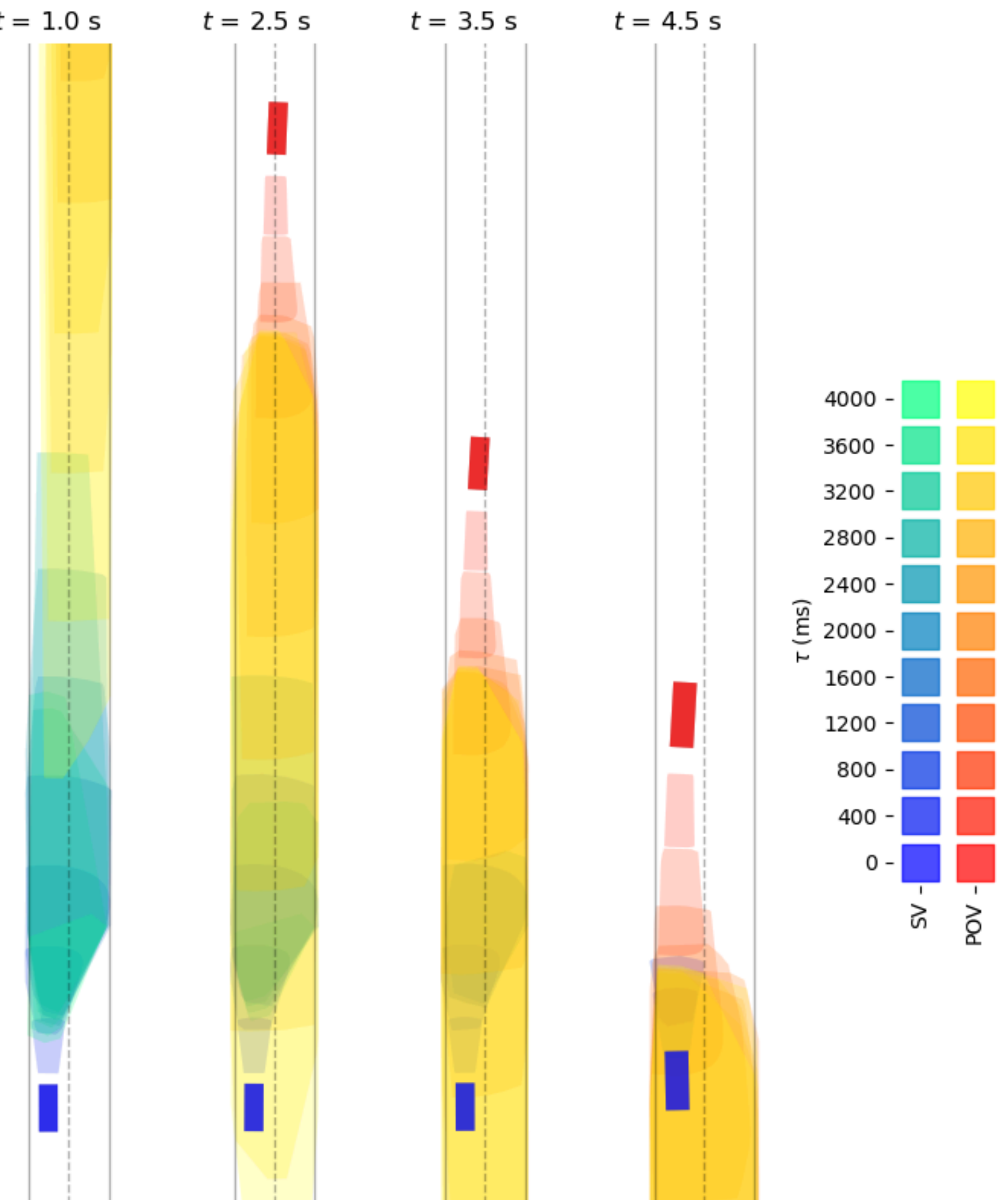

Figure 14. Illustration of reachability modeling for one participant in the medium incursion level. Otherwise same as Figure 13.

Finally, Figure 15 shows an example output from reachability analysis for a participant in the shallow incursion scenario. This scenario is similar to the steep scenario in the sense that the SV initially has space in its lane, then loses its drivable area, and finally regains an affordance near the end of the interaction. In this case, the affordance opens up when the POV enters into a straight path half way into the SV's lane. This is due to the current assumption that the POV will not swerve further to the SV's left than its current heading (i.e., zero assumed lateral acceleration to the POV's right, see Table 1), which corresponds to a belief in the SV driver that the POV will likely continue straight or turn back into its own lane, but not swerve further towards the shoulder. That human drivers would entertain such a belief about the POVs future trajectory could be generally motivated by the established phenomenon of *continuation expectancy*, which is a largely automated tendency of humans to expect the continuation of the perceived events or states of affairs in the immediate future (Näätänen & Summala, 1976) plus the assertion that human drivers are biased towards believing that the POV, after the incursion, will turn back into its own lane. Without this assumption, the affordance typically does not open up until significantly later, demonstrating how the escape affordances depend critically upon the driver's beliefs (behavior predictions) and the uncertainty associated with those beliefs.

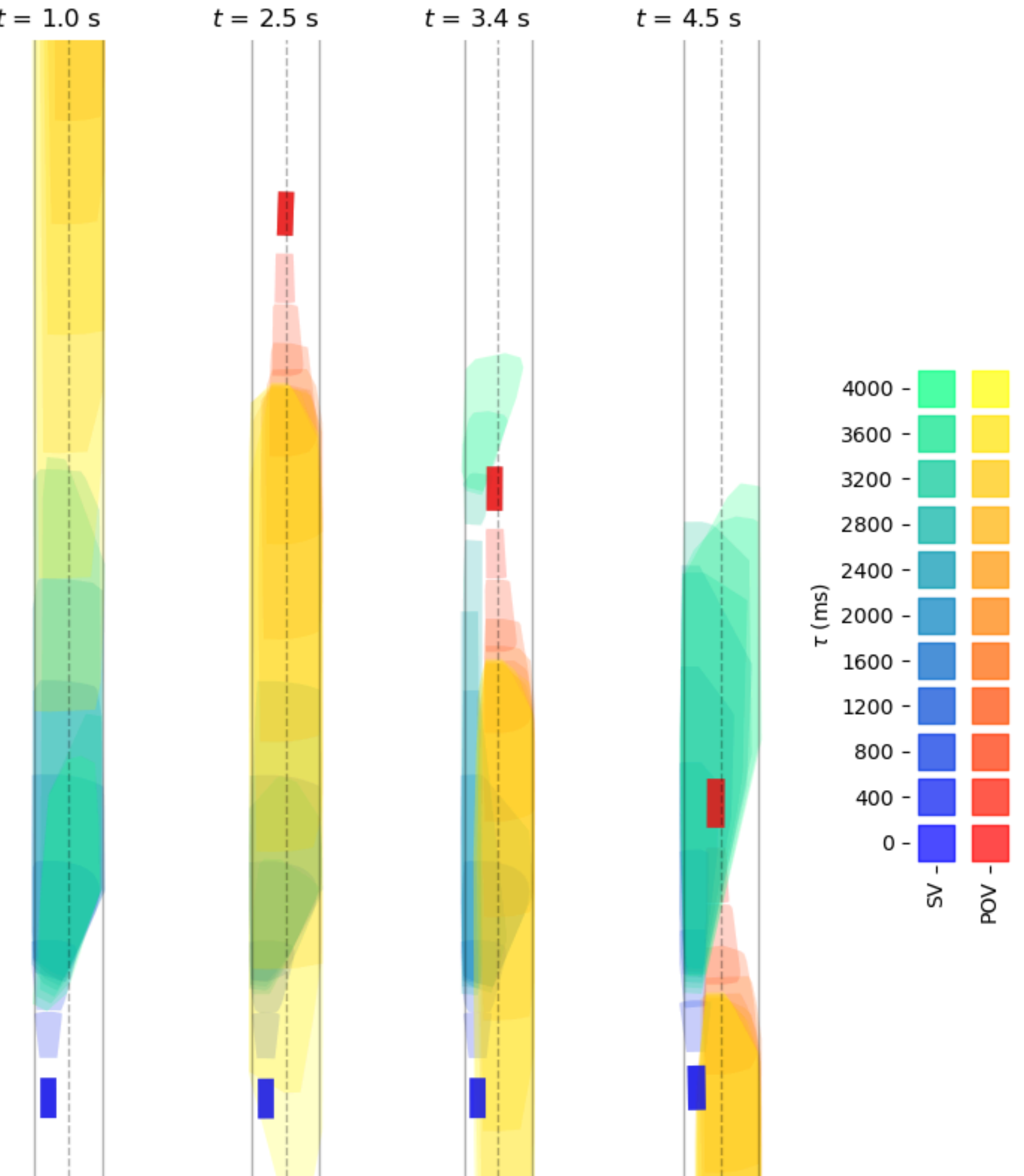

Figure 15. Illustration of reachability modeling for one participant in the shallow incursion level. Otherwise same as Figure 13.

Figure 16 shows an aggregate analysis of the evolution of the drivable area over time, for all participants in the three scenarios. As can be observed, a drivable area opens up earliest on average in the shallow scenario, while drivable areas in the steep and medium scenarios do not appear until almost 1-1.5 seconds later (at the earliest). Given that the participants' beliefs about the POV's future path are at least roughly aligned with the current kinematic assumptions in Table 1, this offers a potential explanation for the response time pattern reported above (Figure 11), where the response times in the shallow scenario were significantly faster than the steep scenario, as further discussed below.

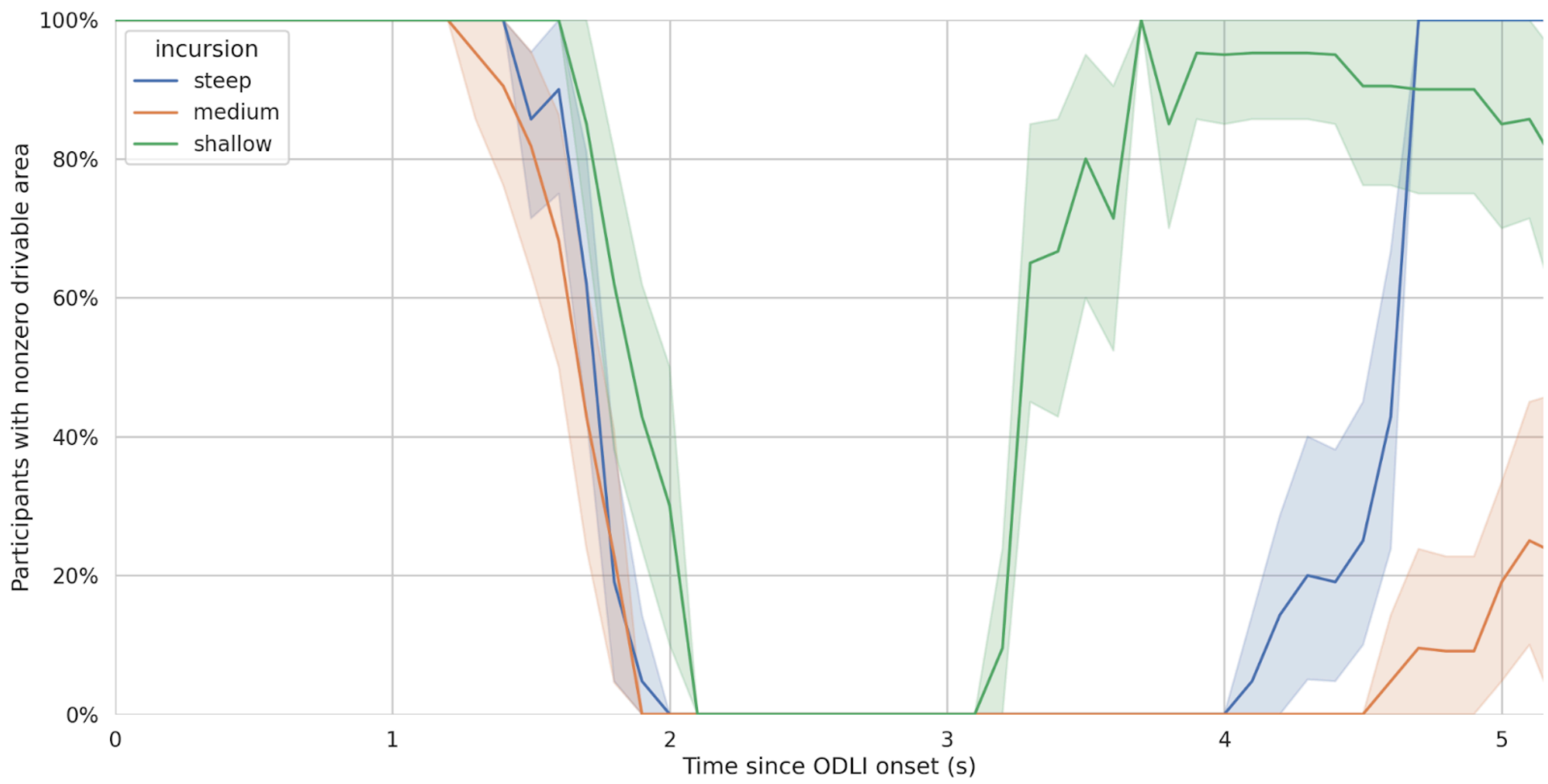

Figure 16. Aggregate prevalence of having a drivable area over time. Solid lines indicate the fraction of participants that have a drivable area at each time step in the analysis window; shaded regions indicate the bootstrapped 95% confidence interval.

# Discussion

The results from the present study generally confirm the findings from previous studies (Hu & and Li, 2015; Lechner & Malaterre, 1988; Weber et al., 2017) that human driver evasive maneuvering behavior is strongly determined by variations in scenario kinematics. While the previous studies focused on lateral incursions in SCP scenarios, the present study additionally showed that this also holds for opposite direction lateral (ODLI) incursions. The results showed that the great majority of participants avoided collision in the steep and shallow incursion scenarios by consistently steering right (opposite direction) and left (same direction) respectively. However, in the more ambiguous medium incursion scenario, the participants' evasive strategies varied strongly and the great majority of the participants crashed, even if the urgency (TTC at the start of the incursion) was the same as in the other two scenarios, and a collision was kinematically avoidable in hindsight by steering to the right without braking. This leads to the important conclusion that *human drivers are able to find and utilize available escape paths, here conceptualized as affordances, when they exist and are unambiguously specified by the scenario (see Figure 13 and 15). However, in scenarios that lack strong, unambiguous, affordances, due to uncertainty about the future trajectory of the other road user, drivers struggle to avoid collision, even if collision was kinematically avoidable after the fact (see Figure 14).* This highlights the key point, further elaborated below, that the “appropriateness” of an evasive maneuver must be judged without hindsight bias, taking into account the uncertainty about how the situation will play out, and not after the fact when the outcome of the situation is known.

The current study also included a number of novel analyses that shed further light on the factors and mechanisms underlying human driver collision avoidance behavior. First, the analysis of sequences of evasive response actions revealed markedly different response patterns for the three incursion scenarios. In the steep incursion scenario, the participants reliably found an escape path to avoid collision by steering in the opposite direction of the incursion, typically either by braking first and then swerving or by swerving only. This shows that drivers consistently perform opposite direction steering if there is a clearly available affordance for doing so, contrary to suggestions in the literature that opposite direction steering is rarely attempted due to a strong habitual tendency for same direction steering (Hu & Li, 2015; Li et al., 2019). The current finding is supported by the results of Lechner and Malaterre (1988) and Weber et al. (2017), cited above, who also found that their participants reliably opted for opposite direction steering in SCP scenarios with long TTCs, where a gap (an affordance) was expected to materialize behind the incurring POV.

By contrast, in the medium incursion scenario, the participants' response sequences were much more diverse and often involved initial steering to the left (same direction) followed by a reversal of steering towards the right (opposite direction), typically combined with braking. This may be interpreted as support for an inherent habitual bias for same direction swerving based on a belief that the POV would turn back into its lane, which was altered as the scenario unfolded (and the POV did not turn back). Additionally, early same direction swerving could function as a form of communication of intent to the POV, creating a stronger affordance for the POV driver to steer back to its lane. However, as shown by the sequence plot for the medium incursion scenario (Figure 9), a significant portion of the participants also opted for initial steering in the opposite direction. It should also be noted that the same direction steering may have been successful if the POV had indeed turned back into its lane, in which case the currently successful strategy (adopted by three participants) to swerve in the opposite direction would likely have led to collision, again emphasizing the importance of avoiding hindsight bias.

Finally, for the shallow incursion scenario, the participants reliably avoided collision by steering left (same direction), the great majority without braking. Again, this shows that same direction steering is not necessarily an “irrational” habitual response (Hu and Li, 2015; Li et al, 2019) implying “swerving into danger” (Weber et al., 2017) but rather an efficient collision avoidance strategy if there is a strong affordance for doing so.

The analysis of response time was split into (1) the initial reaction time and (2) the steering or braking response onset time. The former thus corresponds to the "first identifiable action" defined by Lechner and Malaterre (1991) and the latter corresponds to their concept of "first active input". There was no effect of incursion level on the initial reaction time, in line with the results of Lechner and Malaterre (1991) for "first identifiable action". This lack of effect is likely due to the initial reaction being a preparatory, "reflexive", response to the event before the driver has been able to fully assess the situation (Lechner and Malaterre, 1991). Hence, the initial reaction would not be expected to be much influenced by the specific scenario kinematics. By contrast, there was a significant effect of incursion level on response time such that the steep incursion had the longest response time, the shallow incursion scenario the shortest response time and while the response times for the medium incursion scenario fell in between. A further separate analysis of braking and steering response time revealed that this effect of incursion level was mainly driven by differences in steering response time. This result is generally in line with Lechner and Malaterre (1991) who also found a general effect of scenario kinematics on response time and more specifically consistent with the results of Li et al., (2019) for the ODLI scenario, where they found a significantly longer response time for opposite steering compared to same direction steering (although, in their case, all within a single ODLI scenario, without kinematic variations). In the present study, based on inspection of video from the scenarios, the longer response time for the steep scenario (mainly involving opposite direction steering) could at least be partly explained by the fact that a significant portion of the participants primarily avoided collision by initial braking, and only steered slightly to the right at the very end of the scenario, which hence leads to prolonged steering response times. However, as suggested above, the reachability analysis also suggests that the faster response times for the shallow scenario (mainly involving same direction steering) may also be due to the fact that (under the current kinematic assumptions) the escape affordance for the shallow scenario materializes earlier than for the steep and medium scenarios, where the times where the affordance first appears in Figure 16 for the three incursion scenarios shows at least a rough similarity to the response times in Figure 11.

It can also be noted that the response times measured in the present study were significantly longer than the response times typically found in existing studies. While our response times are in the 3.5 s - 4 s range, Li et al. (2019) in a similar ODLI scenario as ours found steering response times of 1.41 s (for same direction steering) and 3.38 s (for opposite direction steering), and response times from the SCP scenarios are even shorter, on the order of 1 s (Lechner and Malaterre, 1991). This may be explained by the rather gradual development of the incursion in our ODLI scenarios while, by contrast, in Li et al. (2019), the POV started from a stopped position and instantaneously accelerated up to full speed, which likely created a more abrupt realization of the hazard. This general kinematics-dependence of response times is a well-established phenomenon in the driver behavior literature (Markkula et al., 2016; Engström et al., 2024) and has been explained and modelled based on the fact that different scenario kinematics are associated with different times needed to accumulate sufficient evidence for the need to execute an evasive response (Markkula et al., 2016; Engström et al., 2024). In the present study, the evidence developed rather slowly while in Li et al. (2019) it likely developed faster, thus triggering earlier responses. However, the present results further suggest that the accumulation of "evidence" for the need to execute an evasive response may be related to the reduction in perceived uncertainty, yielding clearer affordances, as the scenario unfolds, as shown by the current reachability analysis visualized in Figures 13-15. This idea is supported by recent modeling work by Schumann et al. (2025), partly based on the present data, but needs to be further investigated experimentally.

One limitation of the current study, shared with most of the existing work reviewed above, concerns the ecological validity of behavioral results obtained in driving simulators. In particular, the fact that simulated scenarios do not represent real world risks may play into drivers' evasive maneuvering decisions (Engström and Aust, 2011). In addition, it should be noted that the driving simulator used in the present study was at the lower end of the fidelity range. It should not be expected that the results presented here would align exactly with how drivers would behave in the real world for the same situations. However, simulator studies often provide at least relative validity (Wynne et al., 2019), which in our case would mean that the relative effects of ODLI kinematics might be correctly captured, even if the absolute behaviour is not. With increasing amounts of high-fidelity naturalistic driving data it will be possible to validate driving simulator-based findings on human collision avoidance behavior in more ecologically valid settings, and this is clearly an important avenue for future research.

With these caveats in mind, the present results paint a relatively clear picture where the collision avoidance strategies of human drivers can be understood as driven by affordances, representing available escape paths, which are strongly determined by the perceived uncertainty of how the situation will play out in the near future. Thus, when facing an urgent traffic conflict, drivers are continuously "looking for an out" (an escape affordance) taking into account their perceived uncertainty about the future movement of other road users. The current reachability analysis further illustrated how this notion of affordances can be operationalized and visualized as continuously unfolding kinematically reachable future states that are collision free under "reasonable worst case assumptions" about other road users' future behavior. These affordances can thus be seen as opportunities for evasive action that can be further used as the basis for planning and selecting such actions (see Schumann et al., 2025).

The present results suggest that a key factor determining the perception of affordances for evasive action is the kinematics of the situation and the associated uncertainty. However, as discussed above, it is possible that affordances are also shaped by prior beliefs of human drivers, such as continuation expectancy (Näätänen & Summala, 1976) and an assumption that other drivers are goal-driven and rational. In the ODLI scenario, this would suggest a bias towards the prediction that the other driver is more likely to return to its original lane than continuing across (as in the current steep and medium incursion scenarios). However, the present study was not set up to disentangle the role of such prior beliefs from the role of  scenario kinematics per se, and this is an interesting topic for further experimental research.

It should be noted that the conceptualization of affordance suggested here differs in some fundamental ways from the traditional way the concept is used in ecological psychology (e.g., Gibson, 2014 / 1979; Chemero, 2003, 2009).  While the precise definition of affordances is debated even within ecological psychology (see e.g. Chemero, 2009), the traditional affordance concept rests on the ecological psychology tenet of *direct perception*, which suggests that information needed to guide action is directly specified by the sensory array based, for example, on invariant sensory patterns such as optical flow, without the need for inference based on mental representations. Thus, on this view, affordances are directly picked up from the environment through perception, bypassing cognition. Chemero (2009) summarizes this view as follows: "*If perception is direct, no information is added in the mind; if perception also guides behavior, the environment must contain sufficient information for the animal to guide its behavior. That is, the environment must contain information that specifies opportunities for behavior. In other words, the environment must contain information that specifies affordances.*" (p. 106). This, however, leaves little room for accounting for an organism's uncertainty about the perceived environment and, hence, this traditional notion of affordances seems to be of limited utility as a basis for conceptualizing available escape paths in collision avoidance. For example, in the oncoming incursions in the current study, there is initially no clearly available escape path offered to the driver for steering left or right, given that the driver is strongly uncertain about where the POV will go in the next few seconds. In this case, the affordance for an evasive maneuver only materializes as the scenario unfolds and the uncertainty about the future location of the oncoming vehicle is reduced. In other words, reliable information about whether to steer left or right cannot simply be picked up directly from the optical array, because the driver is uncertain about how the situation is going to play out.

Thus, to be useful for present purposes, a conceptualization of affordance applicable to collision avoidance necessarily needs to factor in uncertainty in the driver's beliefs about how the situation will evolve but this contradicts the fundamental assumption in ecological psychology that perception of affordances is direct. A further limitation of the classical notion of affordance, when applied to collision avoidance, is that it does not specify what actions are *desirable* to the driver, in this case to avoid collision (and, if possible, while avoiding too harsh braking and staying on the road). At the same time, as we hoped to have demonstrated here, the general notion of affordances as agent-relative opportunities for actions (here escape actions) is very valuable for understanding human collision avoidance in driving. Thus, we argue, along with Ramstead (2022), for a broader notion of affordance that may be at odds with the classical notion but more applicable to practical problems. For our purposes, such a notion could be generally formulated as *available opportunities for agent actions expected to lead to desirable outcomes, given the agent's beliefs about possible future development of the situation, and the uncertainty associated with those beliefs*. As demonstrated in this paper, one way to operationalize this broader notion of affordance is in terms of reachable sets. Alternatively, as suggested by Ramstead (2022) and Friston (2022), in the framework of active inference, affordances may be modelled in terms of the negative expected free energy of a given action policy (or plan). This latter option is explored in Schumann et al. (2025)) which demonstrated that an active inference-based computational driver model was able to reproduce the key behavioral and collision outcome results from the current study.

Related to this work, the present results have several implications for the development of computational models of collision avoidance behavior. First, in order to capture sequences of evasive actions, especially in ambiguous, uncertain scenarios such as our medium incursion scenario, models need to represent closed loop control, allowing for incremental replanning of evasive actions as the scenario unfolds. Second, when deciding on an evasive maneuver, models need to account for the uncertainty associated with different potential escape paths (affordances). Third, models need to account for the dependency of response timing on scenario kinematics. For example, a fixed response time model would not work as the 4 s response times exhibited here relative to the incursion start are far beyond any “canonical” response time suggested in the literature (e.g., Green, 2000). In addition, as discussed by Markkula et al. (2016) and Engström et al. (2024), it is often challenging to precisely define the stimulus onset in gradually developing scenarios such as the current ODLI incursion. Thus, collision avoidance behavior is better thought of in terms of a decision making process based on gradual accumulation of evidence and involving the continuous evaluation of potential affordances as the scenario unfolds, resulting in a sequence of response actions, as opposed to a “single shot” action taking place after a given response time.

Moving forward, computational models of human collision avoidance behavior could make detailed predictions of human evasive trajectories in scenarios with given kinematics which can then be evaluated against human data in empirical studies (both experimental driving simulator studies and naturalistic driving studies). In line with Carvalho and Lampinen (2025) we envision that such iterative model development and empirical testing will incrementally lead to a more detailed understanding, and refined generalizable computational models, of human collision avoidance behavior in naturalistic settings.

# CRediT authorship contribution statement

**Leif Johnson**: Conceptualization, Data curation, Formal analysis, Methodology, Visualization, Writing – original draft, Writing – review & editing. **Johan Engström**: Conceptualization, Methodology, Project administration, Supervision, Writing – original draft, Writing – review & editing. **Aravinda Ramakrishnan Srinivasan**: Conceptualization, Data curation, Formal analysis, Investigation, Methodology, Visualization, Writing – review & editing. **Ibrahim Özturk**: Data curation, Investigation, Writing – review & editing. **Gustav Markkula**: Conceptualization, Methodology, Project administration, Supervision, Writing – original draft, Writing – review & editing.

# Conflict of Interest

LJ and JE were employed by Waymo LLC and conducted the research without any external funding from third-parties. Techniques discussed in this paper may be described in U.S. Patent Application No. 19/092,578. The remaining authors declare that the research was conducted in the absence of any commercial or financial relationships that could be construed as a potential conflict of interest.